\documentclass[11pt,a4paper]{article}
\usepackage{jcappub}
\usepackage{bm}
\usepackage{amsmath}
\usepackage{enumitem}

\def\dd{\mathrm{d}}

\def\mcP{\mathcal{P}}

\def\Mpl{M_{\rm Pl}}
\def\GeV{{\rm GeV}}

\def\osc{{\rm osc}}

\def\0{{(0)}}
\def\sig0{\dot{\sigma}_0}

\def\ph0{\dot{\phi}_0}

\title{
Tensor Spectra Templates for
Axion-Gauge Fields Dynamics
during Inflation
}
\author[a,b]{Tomohiro Fujita,}
\author[c,d]{Evangelos I. Sfakianakis}
\author[e]{and Maresuke Shiraishi}

\affiliation[a]{Department of Physics, Kyoto University, Kyoto, 606-8502, Japan}
\affiliation[b]{D\'epartment de Physique Th\'eorique and Center for Astroparticle Physics, \\
Universit\'e de Gen\`eve, Quai E.
Ansermet 24, CH-1211 Gen\`eve 4, Switzerland}
\affiliation[c]{Nikhef, Science Park 105, 1098 XG Amsterdam, The Netherlands}
\affiliation[d]{Lorentz Institute for Theoretical Physics, Leiden University, \\
2333CA Leiden, The Netherlands}
\affiliation[e]{Department of General Education, National Institute of Technology,
Kagawa College, 355 Chokushi-cho, Takamatsu, Kagawa 761-8058, Japan}

\emailAdd{t.fujita@tap.scphys.kyoto-u.ac.jp}
\emailAdd{e.sfakianakis@nikhef.nl}
\emailAdd{shiraishi-m@t.kagawa-nct.ac.jp}

\abstract{$SU(2)$ gauge fields can generate large gravitational waves during inflation, if they are coupled to an axion which can be either the inflaton or a spectator field. 
The shape of the produced tensor power spectrum $\mcP_h$ depends on the form of the axion potential.
We derive analytic expressions and provide general templates for $\mcP_h$ 
for various types of the spectator axion potential.
Furthermore, we explore the detectability of the oscillatory feature, which is present in $\mcP_h$ in the case of an axion monodromy model, by possible future CMB B-mode polarization observations.
}

\arxivnumber{1812.03667}

\begin{document}

\begin{flushright}
Nikhef 2018-063
\end{flushright}

\maketitle

%
%
%

\section{Introduction}

Inflation remains the leading paradigm for the early universe, elegantly explaining the observed flatness, homogeneity and isotropy of the universe, as well as the absence of monopoles \cite{Guth:1980zm, Linde:1981mu}. Perhaps more importantly, it provides a framework for computing primordial fluctuations that are manifested as CMB anisotropies and seed the evolution of Large Scale Structure (LSS). The predictions of the various inflationary models have been compared to observations and after the latest {\it Planck} results \cite{Akrami:2018odb} a large number of inflationary models remains viable.

Despite the success of the inflationary paradigm, the underlying particle physics model driving inflation remains unknown, as does its connection to the subsequent particle content of the Universe\footnote{Using the Higgs field as the inflaton, which requires a large non-minimal coupling to gravity,  leads to a model that provides an unambiguous connection to the Standard Model \cite{Bezrukov:2007ep, GarciaBellido:2008ab, Bezrukov:2008ut, Sfakianakis:2018lzf}. }. A generic prediction of  inflation that has not been yet verified is the presence of primordial gravitational waves. Gravitational waves are produced in all inflationary models due to the presence of quantum fluctuations in the underlying space-time, which are stretched to cosmological scales by the quasi-de-Sitter expansion of the Universe. An ambitious experimental effort is planned to detect the effect of primordial gravitational waves on the CMB through B-mode polarization. These experiments
such as the LiteBIRD satellite mission~\cite{Matsumura:2013aja} and the CMB Stage 4 initiative~\cite{Abazajian:2016yjj} are expected to reach a sensitivity able to detect a tensor to scalar ratio as low as $r\sim 10^{-3}$ (see for example Ref.~\cite{Abazajian:2013vfg, Errard:2015cxa}). 

The amplitude of gravitational waves in most inflationary models is ultimately characterized by the inflationary energy scale, which is related to $r$ as $V_{\rm infl}^{1/4} \approx 10^{16} (r/0.01)^{1/4}\, $GeV on the CMB scales.
It is also directly linked to the inflaton's excursion in field-space through the Lyth bound \cite{Lyth:1996im} in simple models of inflation, $\Delta\phi\gtrsim 0.1\Mpl(r/0.01)^{1/2}$.
Concurrently to the experimental efforts to detect primordial gravitational waves, an ongoing model building effort has been exploring ways to evade the Lyth bound and decouple the tensor mode amplitude from the scale of inflation. A scenario which offers an alternative way to generate gravitational waves during inflation is 
Chromo-Natural Inflation (CNI). CNI was inspired by natural inflation \cite{Adams:1992bn, Freese:1990rb}, where a pseudo-scalar axion field plays the role of the inflaton and its action is protected by a softly broken shift symmetry. By coupling the axion to an $SU(2)$ field through a Chern-Simons term $\phi F\tilde F$, a new source of friction is introduced to the axion dynamics, leading to a new slow-roll attractor, even if the axion potential was initially too steep \cite{Adshead:2013nka, Adshead:2013qp, Adshead:2012kp, Maleknejad:2016qjz,Papageorgiou:2018rfx}
(see also a related model where the axion is integrated out \cite{Maleknejad:2011jw,Maleknejad:2011sq,Maleknejad:2012fw}).

As a reminiscent feature of $U(1)$ fields coupled through a Chern-Simons term \cite{ArmendarizPicon:2007iv, Anber:2009ua,Adshead:2015pva,Fujita:2015iga, Namba:2015gja,Adshead:2016iae}, the tensor modes of the $SU(2)$ sector experience an instability and are exponentially amplified. This only occurs for one of the two polarizations. The amplified $SU(2)$ tensors seed gravitational waves, which are also chiral.
The analysis of the spectral index is rather complicated and it was shown that the original version of CNI, where the axion potential was taken to be of $V(\chi) \propto 1 -\cos(\chi/f)$, is not compatible with CMB observations \cite{Adshead:2013nka}, because the scalar spectral tilt $n_s$ was too small for observationally allowed values of $r$. This can be remedied if the $SU(2)$ gauge symmetry is spontaneously broken. The resulting model of Higgsed Chromo-Natural Inflation was studied in Ref.~\cite{Adshead:2016omu} and was shown to provide observables within the Planck-allowed region for certain parts of parameter space, while evading the Lyth bound and generating observable gravitational waves at a lower inflationary scale. Furthermore the resulting tensor spectral tilt $n_T$ generically violated the consistency relation $r=-8n_T$.

A different route was taken in Ref.~\cite{Dimastrogiovanni:2016fuu}, where the Chromo-Natural Inflation action was treated as a spectator sector. Recognizing that the tachyonic instability of the $SU(2)$ tensor modes can source sizeable gravitational waves, even if the energy density in the axion-$SU(2)$ fields is small, an unknown inflaton field can be invoked to generate the observed scalar fluctuations, while the dominant part of the tensor modes is generated by the spectator CNI sector. This further de-couples the inflationary energy scale from the GW amplitude, in principle allowing for very low scale inflation with observable GW's~\cite{Fujita:2017jwq}. A difference of the sourced GW's in this model to the usual vacuum modes is the amount of Non-Gaussianity \cite{Agrawal:2017awz}. Whereas vacuum GW's are very gaussian and respect the parity symmetry, sourced GW's are predicted to exhibit a level of non-Gaussianity and chirality that could be in principle measured by future experiments, such as LiteBIRD~\cite{Thorne:2017jft,Dimastrogiovanni:2018xnn,Fujita:2018vmv}.

Due to their shift symmetry, axions provide an elegant way of addressing the $\eta$ problem of inflation, protecting the flatness of the potential from unknown higher order corrections. This has in part motivated the study of axions in the context of string theory. Several models have been constructed that have different observational consequences \cite{Long:2014dta, McAllister:2008hb, Silverstein:2008sg, McAllister:2014mpa, Kaloper:2011jz, Marchesano:2014mla, Blumenhagen:2014gta, Hebecker:2014eua, Hebecker:2015rya, Cai:2014vua, Baumann:2014nda, Kobayashi:2015aaa, Parameswaran:2016qqq, Kobayashi:2017jeb}, like providing a flat axion potential leading to a lower value for the tensor-to-scalar ratio $r$, or exhibiting oscillatory features. Despite the extensive literature on axions, most CNI and spectator CNI models have relied on the cosine potential (with the exception of Ref.~\cite{Caldwell:2017chz}). In anticipation of future experiments and the possible observation of stochastic primordial gravitational waves, it is will be essential to explore and  classify ways to probe the the physics of axion fields at very high energies. It has been established that the existence of a spectator axion-$SU(2)$ sector during inflation can produce observable gravitational waves. In this work we show that the spectrum of primordial GW's carries information about the underlying axion potential that can be in principle extracted by future experiments. We categorize axion potentials in three main types, based on their morphology for field values relevant for inflation and provide templates for the spectra of the produced  tensor modes.

This paper is organized as follows. In Section~\ref{sec:Model} we set up the model, review the slow-roll analysis and distinguish the three different axion potential types that we will study. The three model types are studied separately in detail. In Section~\ref{sec:powerlaw} we study power-law potentials, as the prototypical example of a model with monotonic convex or concave axion potential. In Section~\ref{sec:generalcosine} we compute the gravitational waves produced in models when the inflaton potential crosses a non-stationary inflection point during inflation. 
The axion monodromy potential, containing an oscillatory term, is studied in Section~\ref{sec:monodromy}, where special attention is given to the departure from the usual slow-roll results found in the literature for spectator axion-$SU(2)$ models. Section~\ref{sec:CMB} examines the observability of the computed tensor spectra by future CMB missions, with special emphasis on the case of a modulated axion monodromy potential.
  We summarize our work and offer our conclusions in Section~\ref{sec:summary}.

\section{Model}
\label{sec:Model}

Several constructions originating in string theory \cite{Long:2014dta, McAllister:2008hb, Silverstein:2008sg, McAllister:2014mpa, Kaloper:2011jz, Marchesano:2014mla, Blumenhagen:2014gta, Hebecker:2014eua, Hebecker:2015rya, Cai:2014vua, Baumann:2014nda, Kobayashi:2015aaa, Parameswaran:2016qqq, Kobayashi:2017jeb} have been proposed for generating an axion potential. While we do not attempt to provide an exhaustive description of all possible axion models, we can define three main phenomenological types, which can be used to classify most cases of interest, in the context of spectator axion-gauge dynamics during inflation. Table~\ref{tab:Types} shows the three main potential types that we will consider, along with an example of each type. Type\ I describes potentials that are monotonic and remain convex or concave in the whole range of values that the spectator axion field acquires during inflation.   Type\ II describes potentials in which the oscillatory term is dominant and the axion probes a single non-stationary inflection point $U'(\chi)=0$ of the potential during its evolution during inflation. The prototypical cosine potential first associated with natural inflation  \cite{Freese:1990rb, Adams:1992bn} falls under this category for $p=2$. This is also the most studied axion potential in Chromo-Natural inflation \cite{Adshead:2012kp, Adshead:2013qp, Adshead:2013nka} along with its Higgsed \cite{Adshead:2016omu} and spectator variants~\cite{Dimastrogiovanni:2016fuu,Fujita:2017jwq,Agrawal:2017awz,Thorne:2017jft,Dimastrogiovanni:2018xnn,Fujita:2018vmv}. Finally Type\ III describes an axion monodromy potential, in which the axion probes multiple periods of the modulated potential during inflation.

Table~\ref{tab:Types} also shows one sample potential form for each of the three types. This can be mostly viewed as a phenomenological choice, and we will discuss the relation of our choice to axion potentials derived from UV theories, like string theory (following the specific forms given in Ref.~\cite{Kobayashi:2015aaa}). Furthermore, we show the resulting form of the Gravitational Wave spectrum for each potential choice\footnote{Specifically for  Type III potentials, the resulting simple oscillatory form of the gravitational wave spectrum is only derived for a linear potential with a small modulation. More complicated potentials will lead to a more complicated GW template. An extensive study of modulated axion potentials  can provide useful templates for GW searches in light of future CMB experiments and can be performed using the techniques discussed in Section~\ref{sec:monodromy}. }.
 We will study each scenario separately and explain in detail how the various Gravitational Wave templates are derived.

 \begin{table}[h]
 \renewcommand{\arraystretch}{2}
 \begin{center}
   \begin{tabular}{ | l | l | l | l | l | l | l | l | l | l | l | l |}
    \hline
      & potential type & sample potential   &  GW template  
     \\ \hline
    Type\ I & convex / concave & $U(\chi)\propto \chi^p$  & $\mcP_h^{\rm (s)}(k)\propto \left(\frac{k}{k_*}\right)^{n_T} $
     \\ \hline
          Type\ II & one inflection point  & $U(\chi)\propto \left [ 1-\cos\left ({\chi\over f} \right ) \right ] ^{p\over 2}$  & $\mcP_h^{(s)} \propto \exp \left [- {{ \ln^2\left ( {k/ k_*}\right )\over 2\sigma_h^2} }\right ]$ 
              \\ \hline
     Type\ III & axion monodromy & $U(\chi)\propto \chi^p+ \delta \cos(\nu \chi) $  & $\mcP_h^{(s)} \propto 1+A \sin\left [C \ln({k\over k_*})+\theta\right]
$  
    \\
 & (modulated) & & \hspace{ 1.6cm} for  $p=1 ~\&~ {A}\ll 1$
    \\ \hline
     \end{tabular}
     \end{center}
     \renewcommand{\arraystretch}{1}
 \caption{
General types of axion potentials transversed by the spectator field during inflation, along with the template for the resulting gravitational wave spectra.
For our purposes axion potentials can be categorized by the number of non-stationary inflection points $U''(\chi_*)=0, \, U'(\chi_*)\ne0$ in the relevant field range. Potentials of Type I, II and III have zero, only one and multiple inflection points, respectively.
  } 
 \label{tab:Types}
\end{table}%

\subsection{Slow Roll background}
\label{Slow Roll background}

Let us consider the following action~\cite{Dimastrogiovanni:2016fuu} 
\begin{align}
S=\int \dd^4 x \sqrt{-g}
&\left[
\frac{1}{2}\Mpl^2 R-\frac{1}{2}(\partial \varphi)^2-V(\varphi)
\right.\notag\\
&\quad\left.
-\frac{1}{2}(\partial \chi)^2-U(\chi)-\frac{1}{4}F_{\mu\nu}^a F^{a\mu\nu}
+\frac{\lambda}{4f}\chi F_{\mu\nu}^a \tilde{F}^{a\mu\nu}
\right],
\label{action}
\end{align}
where $\Mpl$ is the reduced planck mass, $R$ is Ricci scalar, $\varphi$ denotes the inflaton with the potential $V(\varphi)$, $\chi$ is a spectator (non-inflaton) pseudo-scalar field (axion) with the potential $U(\chi)$, $F_{\mu\nu}^a\equiv \partial_\mu A_\nu^a- \partial_\nu A_\mu^a-g \epsilon^{abs}A_\mu^b A_\nu^c$ is the field strength of a SU(2) gauge field $A_\mu^a$, and $\tilde{F}^{a\mu\nu}\equiv \epsilon^{\mu\nu\rho\sigma}F^a_{\rho\sigma}/(2\sqrt{-g})$ is its dual. 
The parameters $g$ and $\lambda$ are dimensionless coupling constants, while $f$ is the axion decay constant (of the field $\chi$) and has dimensions of mass.
In previous works the cosine type potential of the axion field has been investigated\footnote{The sign of the cosine term can be altered by a simple constant shift of the axion field value $\chi\to \chi+\pi f$ and thus has no real physical meaning.},
\begin{equation}
U(\chi) =\mu^4 \left[1 - \cos \left(\frac{\chi}{f}\right)\right] \, ,
\label{Cos potential}
\end{equation}
which was the originally proposed potential in the context of natural inflation \cite{Freese:1990rb, Adams:1992bn}.
In this paper, we extend previous work by considering more general types of potentials, as shown in Table~\ref{tab:Types}.

{Without loss of generality, the axion background field is assumed to have a negative initial value, $\chi_{\rm in}<0$, and roll toward its potential minimum at $\chi=0$, developing a positive velocity $\dot \chi>0$.
}
We also consider that the gauge fields $A_\mu^a$ are in the classical configuration,
\begin{equation}
A_0^a=0,
\qquad
A_i^a =\delta^a_i a(t) Q(t),
\label{eq:Abackground}
\end{equation}
where $a(t)$ is the scale factor. The non-trivial form of the gauge field background given in Eq.~\eqref{eq:Abackground} has been shown to be stable in the context of Chromo-Natural Inflation~\cite{Maleknejad:2013npa,Domcke:2018rvv}.

From the action of Eq.~\eqref{action} we derive 
the equations of motion for the background axion and the gauge field,
\begin{align}
\ddot{\chi}+3H\dot{\chi}+U'(\chi)
&=-\frac{3g\lambda}{f}Q^2 \left(\dot{Q}+HQ\right),
\label{Full EoM chi}
\\
\ddot{Q}+3H\dot{Q} +\left(\dot{H}+2H^2\right)Q 
+2g^2 Q^3 &=\frac{g\lambda}{f} Q^2 \dot{\chi},
\label{Full EoM Q}
\end{align}
where an overdot denotes the derivative with respective to cosmic time $\dot f(t) = df/dt$, and $H\equiv \dot{a}/a$ is the Hubble parameter.


It has been shown that if the coupling between the axion and the $SU(2)$ sector is strong,
\begin{equation}
\Lambda\equiv \frac{\lambda Q}{f} \gg 1.
\label{Lambda def}
\end{equation}
and the effective mass of the $SU(2)$ field due to its background configuration and self-coupling is significant%
\footnote{It has been shown that for $m_Q<\sqrt{2}$ the scalar perturbations
of the axion-$SU(2)$ system have a fatal instability in the slow-roll regime~\cite{Dimastrogiovanni:2012ew}.
We therefore only consider parameter combinations leading to $m_Q>\sqrt{2}$.},
\begin{equation}
m_Q\equiv \frac{gQ}{H}\gtrsim1,
\label{mQ definition}
\end{equation}
the coupled system enters the slow-roll regime~\cite{Adshead:2012kp}. We can then
 approximate the background equations, Eqs.~\eqref{Full EoM chi} and \eqref{Full EoM Q}, by
\begin{align}
U'(\chi)
&\simeq-\frac{3g\lambda}{f}HQ^3,
\label{chi slow-roll Eq}
\\
2H^2 Q 
+2g^2 Q^3 &\simeq\frac{g\lambda}{f} Q^2 \dot{\chi},
\label{Q slow-roll Eq}
\end{align}
where we have dropped all terms with time derivatives except for the right-hand-side of Eq.~\eqref{Full EoM Q} which transfers part of the axion kinetic energy into the background gauge field.
In this regime, the  potential force and the additional friction from the gauge field (not Hubble friction) are balanced, and $Q(t)$ acquires an almost constant value supported by the kinetic motion of the axion. 
Introducing the dimensionless parameter
\begin{equation}
\xi\equiv \frac{\lambda \dot{\chi}}{2fH},
\label{eq:xi_def}
\end{equation}
Eq.~\eqref{Q slow-roll Eq} is recast as
\begin{equation}
\xi \simeq m_Q+m_Q^{-1}.
\label{xi mQ}
\end{equation}
On the other hand, Eq.~\eqref{chi slow-roll Eq} yields $Q$ 
and $m_Q$ as
\begin{equation}
Q \simeq \left(\frac{-f U'}{3\lambda g H} \right)^{\frac{1}{3}},
\qquad
m_Q\simeq\left(\frac{-g^2f U'}{3\lambda  H^4} \right)^{\frac{1}{3}} .
\label{Q minimum}
\end{equation}
%

\subsection{Gravitational Wave Power Spectrum}

{We introduce the fluctuations of the gauge field $\delta A_\mu^a$ around the background value of Eq.~\eqref{eq:Abackground} as~\cite{Agrawal:2017awz}
\begin{equation}
\delta A_i^a=t_{ai}+\cdots,\qquad
t_{ii}=\partial_{i}t_{ij}=\partial_{j}t_{ij}=0,
\label{eq:deltaAmu}
\end{equation}
where the scalar and vector fluctuations in $\delta A_0^a$ and $\delta A_i^a$ are ignored.
We will instead focus on the transverse-traceless  tensor modes of the $SU(2)$ sector, $t_{ij}$, which can be decomposed into $t_L$ and $t_R$ in the left / right helicity basis (see Eq.~\eqref{mode decomposition} for detail). 
}

In this spectator axion-$SU(2)$ model, the tensorial perturbation of the $SU(2)$ gauge field $t_{L,R}$ undergoes a transient instability around horizon crossing for one of the two polarizations and is substantially amplified. The dominant polarization depends on the sign of $\xi$, defined in Eq.~\eqref{eq:xi_def} and is thus  controlled by the direction in which the axion field $\chi$ is rolling. 
{With our current choice of $\chi<0$ and $\dot{\chi}>0$, the right-handed mode is amplified.}
The amplified mode $t_{R}$ acts as a source term in the equation of motion for gravitational waves with the same polarization.
As derived in Ref.~\cite{Dimastrogiovanni:2016fuu},
the  power spectrum of the sourced gravitational waves
is given by
\begin{equation}
\mcP_h^{(s)} = \frac{\epsilon_B H^2}{\pi^2 \Mpl^2} \mathcal{F}^2(m_Q),
\label{Phs slowroll}
\end{equation}
where $\epsilon_B\equiv g^2 Q^4/(H \Mpl)^2$. It is important to note that the time evolutions of $Q, \dot{\chi}$ and $H$ are ignored when deriving this expression. Here $\mathcal{F}(m_Q)$ is a complicated function
whose full expression can be found in Ref.~\cite{Dimastrogiovanni:2016fuu}.
It is useful to introduce a fitting formula of $\mathcal{F}(m_Q)$,
\begin{equation}
\mathcal{F}(m_Q)
\equiv \exp[\alpha\,m_Q],
\label{Phs simple}
\end{equation}
with $\alpha \approx 2$.\footnote{If necessary, one can use a more accurate
fitting formula (e.g. $\mathcal{F}\approx \exp[2.4308m_Q-0.0218m_Q^2-0.0064m^3_Q-0.86]$ for $3\le m_Q \le 7$) and a similar calculation can be done for the resulting tensor power spectrum.}
 The parameter $\epsilon_B$ can be rewritten in terms of $m_Q$ as
\begin{equation}
\epsilon_B= \frac{H^2 m_Q^4}{g^2 M_{\rm Pl}^2}\propto m_Q^4
\end{equation}
and the sourced GW power spectrum in turn depends on $m_Q$ as
\begin{equation}
\mcP_h^{(s)} 
\,=\,\frac{m_Q^4 H^4}{\pi^2g^2\Mpl^4}\exp[2\alpha m_Q] 
\,\propto\, m_Q^4 \exp[2\alpha m_Q].
\label{Phm}
\end{equation}
It thus seems enough to compute $m_Q$ as a function of the scale $k$, in order to compute the resulting gravitational wave spectrum. We will see that effects arising from the time evolution of the background quantities $Q$, $\dot \chi$ and $H$ can lead to a slight change in the result of Eq.~\eqref{Phm}.

\section{Power-law Potential}
\label{sec:powerlaw}

We first consider a pure power-law potential,
\begin{equation}
U(\chi) = \mu^4 \left|\frac{\chi}{f}\right|^{p}.
\end{equation}
Its first derivative is given by
\begin{equation}
U'=-p \frac{\mu^4}{f} \left|\frac{\chi}{f}\right|^{p-1},
\end{equation}
and hence we obtain the time evolution of $m_Q$ as
\begin{equation}
m_Q(t) =m_* \left(\frac{\chi(t)}{\chi_*}\right)^{\frac{p-1}{3}},
\qquad
m_*\equiv \left(\frac{pg^2\mu^4}{3\lambda H^4}\right)^{\frac{1}{3}}
\left|\frac{\chi_*}{f}\right|^{\frac{p-1}{3}},
\end{equation}
where $t=t_*$ is the time of horizon crossing of our reference scale $k_*=a(t_*)H$ which can be the CMB pivot scale $k_{\rm pivot}$, and $\chi_*\equiv \chi(t_*)$. Then $\chi(t)$ is expanded as
\begin{equation}
\frac{\chi(t)}{\chi_*}\simeq 1+\frac{H(t-t_*)}{\Delta N},
\end{equation}
with 
\begin{equation}
\Delta N\equiv \frac{\lambda \chi_*}{2f\xi_*},
\qquad \xi_*\equiv \frac{\lambda \dot{\chi}_*}{2fH}.
\label{dN def}
\end{equation}
Thus $m_Q$ can be approximated by
\begin{equation}
m_Q(t) \simeq m_* \left[ 1+ \frac{p-1}{3}\frac{H(t-t_*)}{\Delta N}\right].
\label{power mQ} 
\end{equation}

Using $H(t-t_*)= \ln (k/k_*)$, we can translate the time dependence of $m_Q$ given in Eq.~\eqref{power mQ}
into its $k$ dependence,
\begin{align}
m_Q^{\rm power}(k)\simeq  m_* \left[ 1+ \frac{p-1}{3}\frac{\ln(k/k_*)}{\Delta N}\right].
\label{mQ power}
\end{align}
Substituting Eq.~\eqref{mQ power} into Eq.~\eqref{Phm}, we obtain leading-order result for the tensor power spectrum
\begin{align}
\mcP_h^{\rm power}=\frac{m_*^4 H^4\mathcal{F}^2(m_*)}{\pi^2g^2\Mpl^4} \left(\frac{k}{k_*}\right)^{\frac{2 (p-1)}{3\Delta N}(\alpha m_*+2)},
\label{Ph power}
\end{align}
where we have used $(1+\epsilon)^4 \simeq (1+4\epsilon)\simeq \exp[4\epsilon]$ for the $k$ dependence arising from the prefactor $m_Q^4$ in Eq.~\eqref{Phm}.
Therefore, we obtain a power-law power spectrum with the tensor tilt
\begin{equation}
n_T^{\rm power} = \frac{2 (p-1)}{3\Delta N}(\alpha m_*+2),
\label{nT power}
\end{equation}
in the case of a power-law axion potential.
The value of $n_T^{\rm power}$ depends on $\chi_*$ through $m_*$ and $\Delta N\propto \chi_*$
defined in Eq.~\eqref{dN def}. Hence the tensor tilt is not determined only by the model parameters.
However, its sign is fixed solely by the form of the power law through $p-1$, because $\Delta N\propto \chi_*/\dot{\chi}_*$ is always negative.
Therefore, we find%
\begin{equation}
p=1: {\rm scale\ invariant},
\qquad
p>1: {\rm red\ tilt},
\qquad
p<1: {\rm blue\ tilt}.
\end{equation}
It should be noted that we have ignored the time variation of $H$ and 
disregarded $\mathcal{O}(\epsilon_H)$ contributions to $n_T^{\rm power}$.

%
\begin{figure}[tbp]
  \begin{center}
  \includegraphics[width=110mm]{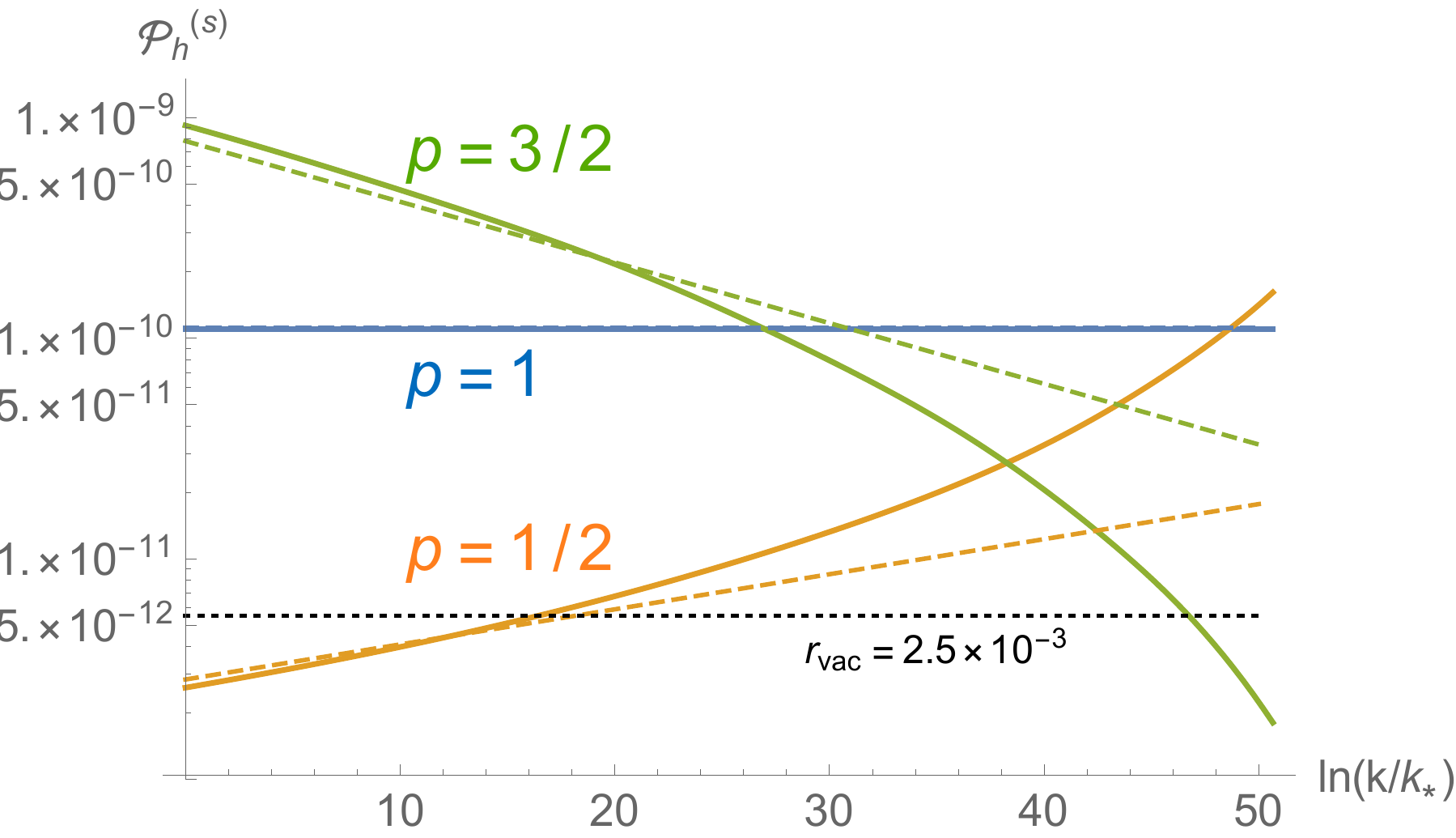}
  \end{center}
  \caption
 {The numerically computed sourced tensor power spectra $\mcP_h^{(s)}$   are shown for the power-law axion potential, $U(\chi)=\mu^4 |\chi/f|^p$ with $p=1$ (blue), $p=1/2$ (yellow) and $p=3/2$ (green). The model parameters are given in Eq.~\eqref{parameter set 1}. The dashed straight lines are the analytically derived $P_{h}^{(s)}$, Eq.~\eqref{Ph power}, with $\alpha=2$ and $\chi_*$ and $m_*$ are evaluated at $k=k_*$. As expected, the sign of the tensor tilt is determined by the power of the axion potential.
}
 \label{three Phs}
\end{figure}
%

In order to check the accuracy of the analytically derived power spectrum $\mcP_h^{\rm power}$
and the resulting spectral tilt $n_T^{\rm power}$ 
given in Eqs.~\eqref{Ph power} and \eqref{nT power} we perform numerical calculations, in which we incorporated the time evolution of the axion field $\chi$ and the resulting background quantities $Q, m_Q$. 
We set the Hubble scale $H(t)$ to be constant, which is an increasingly good approximation for inflationary models with a flat plateau that would produce a small tensor-to-scalar ratio, and 
numerically solve the full equations of motion for the axion field $\chi$ and gauge field background value $Q$, Eqs.~\eqref{Full EoM chi} and \eqref{Full EoM Q}. 

The details of the numerical computation leading to the gravitational wave power spectrum are described in Appendix \ref{Appendix: Numerical Calculation of Tensor Perturbations}.
We use the following parameters
\begin{equation} 
\label{parameter set 1}
\begin{split}
&g=1.11\times10^{-2},~~~~~~~~~\,\lambda=500,~~~~~~~~~~~~~~~~~~~~\  f=4\times10^{16}\GeV,
\\
&H = 1.28\times10^{13}\GeV,\quad \mu = 1.92\times10^{15}\GeV,
\quad \left | \chi_{\rm in} \right |=4\times 10^{16}\GeV,
\end{split}
\end{equation}
which are the same as the ones used in Ref.~\cite{Dimastrogiovanni:2016fuu}.
Fig.~\ref{three Phs} shows the tensor spectra derived using the analytical slow-roll approximation Eq.~\eqref{Ph power} along with the corresponding numerical calculations. For the range $0\le\ln(k/k_*)\lesssim 20$, the tensor power spectra are well approximated by the power-law spectrum, $\mcP_h\propto(k/k_*)^{n_T}$ and their spectral tilt agree with our analytic expressions given in Eq.~\eqref{nT power}.
The small deviations of the amplitudes between the numerical and analytic results at $k=k_*$ originate in the violation of the slow-roll approximation.
Since the tensorial SU(2) perturbation which sources gravitational waves
is amplified slightly before  horizon crossing (see for example Ref.~\cite{Dimastrogiovanni:2016fuu, Adshead:2013nka}), one should use the value of $m_Q$ at that time to evaluate $P_h^{(s)}$. However, we used $m_*$
which is defined as $m_Q(t)$ at the time of horizon crossing for the mode with comoving wavenumber $k_*$.
For $p=1$ the effective mass parameter $m_Q(t)$ stays constant and thus the numerical and analytical values of $\mcP_h^{\rm power}$ are indistinguishable. For smaller values of $p$,  $m_Q(t)$  increases with time and for larger values the opposite happens, $m_Q(t)$ decreases.
 This leads to  the analytic expression overestimating the numerical amplitude of $\mcP_h^{(s)}(k_*)$ for $p=1/2$ and underestimating it for  $3/2$, as shown  in Fig.~\ref{three Phs}.
While the tensor amplitude is exponentially sensitive to $m_Q$
and the deviations are visible, the derived tilt $n_T^{\rm power}$
is not so sensitive and hence it suffices for our purpose.

Fig.~\ref{three Phs} also shows that the discrepancy between the analytically and numerically derived  tensor power spectrum grows for wavelengths much smaller than the one corresponding to $k_*$. This is especially important in the case of a blue-tilted gravitational wave spectrum, which arises for $p<1$. A potential with $p<1$ has a first derivative that diverges at $\chi=0$. This means that our slow-roll solutions will also diverge, signaling a break-down of the slow-roll approximation or of the model itself. Physically, an axion potential that is irregular at small field-values is unrealistic. However, many constructions have been put forward, mostly originating in supergravity or string theory, which generate a power-law axion potential with $p<1$ at large field-values, that is also regular everywhere. In \cite{Amin:2011hj} a simple phenomenological description of these models was used, where the potential was assumed to be quadratic near the minimum and become flatter at large field values
\begin{equation}
U(\chi ) \propto \left (1+{\chi^2 \over M_c^2} \right )^{p/2}-1 \, 
\ \sim 
\left\{
\begin{array}{cc}
|\chi/M_c|^p \quad& (|\chi|\gg M_c) \\[5pt]
\frac{p}{2}\,\chi^2/M_c^2\quad & (|\chi|\ll M_c) \\
\end{array}\right.\,
\end{equation}
where $M_c$ defines the scale that separates the quadratic and flat potential regions. Near the end of inflation, where the axion field is also expected to approach its minimum, the exact form of the potential, including the scale $M_c$ needs to be defined, in order to accurately compute the resulting power in tensor modes  that is generated when the axion field nears the minimum of its potential.

A blue helical gravitational wave spectrum has the capacity to
generate the observed matter-antimatter asymmmetry
  through the standard model lepton-number gravitational anomaly \cite{Alexander:2004us}. 
A recently proposed model of gravitational leptogenesis relies on a modified version of Chromo-Natural Inflation \cite{Caldwell:2017chz}. In order to accurately predict the resulting baryon number several factors need to be taken into account, like the neutrino mass and reheat temperature. However, the prime factor is the power in gravitational waves. Ref.~\cite{Adshead:2017znw} connected the baryon asymmetry to the tensor-to-scalar ratio. 
Spectator models allow for the effective decoupling of the Hubble scale and the tensor-to-scalar ratio. Hence testing the conclusions of Ref.~\cite{Caldwell:2017chz} in the context of the spectator models that we explore here remains an intriguing  open question. 

An important point to note, when attempting to produce a large amount of gravitational waves towards the end of inflation, is the relation between the time when inflation ends and the time when the axion field $\chi$ reaches its minimum. If the latter occurs long after inflation has ended,  the possibility exists that $\chi$ will act as a curvaton field, thereby generating the observed density perturbations (see for example Ref.~\cite{Lyth:2001nq, Bartolo:2002vf, Lyth:2002my} for a description of the curvaton mechanism). Throughout this study, we assume that the axion has relaxed to its minimum during or shortly after inflation, hence it does not spoil the scalar power spectrum that are produced by fluctuations in the inflaton sector.

\section{Generalized Cosine Potential}
\label{sec:generalcosine}

In Ref.~\cite{Caldwell:2017chz} a modified chromo-natural inflation potential was proposed\footnote{The same generalized axion potential form has been recently used in the context of early dark energy \cite{Poulin:2018cxd}.}
\begin{equation}
U(\chi) = \mu^4 \left [1-\cos\left({\chi\over f}\right) \right ]^{p/2}.
\label{eq:Vrobert}
\end{equation}
For $p=2$ Eq.~\eqref{eq:Vrobert} becomes the usual natural inflation cosine potential, Eq.~\eqref{Cos potential} {up to a constant shift $\chi\to \chi+\pi f$}.
It was found in Ref.~\cite{Caldwell:2017chz} that Chromo-Natural inflation with the potential of Eq.~\eqref{eq:Vrobert} can lead to scalar and tensor
power spectra that are compatible with  CMB observations, as well as accommodate  
leptogenesis through the axial-gravitational anomaly
for $1/16\lesssim p\lesssim 1/8$.
The actual computations in Ref.~\cite{Caldwell:2017chz} were performed using the Taylor-expanded potential for $\chi\ll f$, which is the power-law potential described in the present work as ``Type I" and was studied in the previous section. 
In this section, however, we focus on $\chi\sim f$, in order to extract   the characteristic form of the
tensor power spectrum for ``Type II'' potentials, modeled by Eq.~\eqref{eq:Vrobert}.

The potential derivative is
\begin{equation}
U' =  {p\over 2} {\mu^4\over f} \left [1-\cos\left({\chi\over f}\right) \right ]^{p/2}\cot\left({\chi\over 2 f}\right)
\end{equation}
leading to 
\begin{equation}
m_Q (t)  =  \left ( {\frac{-g^2 \mu ^4 p  \left(1-\cos \left({\chi(t)\over f}\right)\right)^{p/2}}{6 \lambda H^4\,\tan \left({\chi(t)\over 2 f}\right) } }\right )^{1/3}
\end{equation}
The derivative exhibits a discontinuity at $\chi =0$ for $p\le1$ and diverges for smaller values of the power $p$. 
In principle Eq.~\eqref{eq:Vrobert} should be modified close to the minimum of the potential at $\chi=0$, for example by the addition of a quadratic term, so as not to induce a diverging spectrum. The potentials described in Section~\ref{sec:trigpotentials} are free from such irregularities. Furthermore, in this Section we are only interested in axion potentials with $p>1$ and thus we will not pursue the pathological behavior arising for $p<1$ any further.

%
\begin{figure}
\centering
\includegraphics[width=100mm]{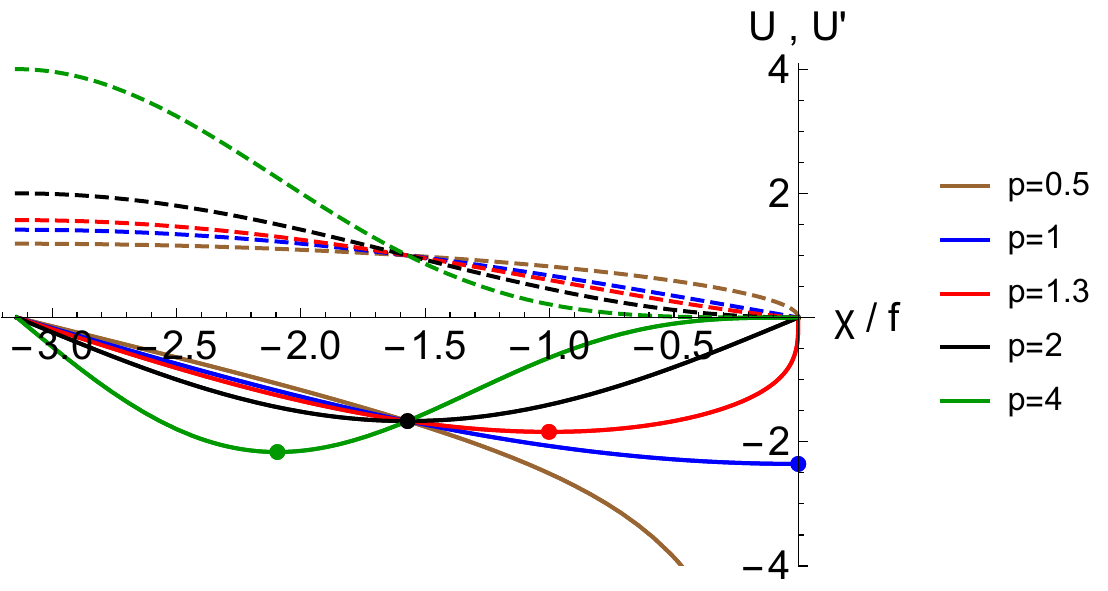}
\label{fig:UandUdotVSchi}
\caption{ 
The potential $U(\chi)$ (dashed) and the derivative $U'(\chi)$ (solid) in arbitrary units for $p=0.5, 1, 1.3, 2, 4$ (brown, blue, red, black and green respectively). The dots correspond to the extremum of  $U'(\chi)$. For $p=0.5$ the potential does not exhibit an inflection point for $0<|\chi/f|<\pi $.
}
\end{figure}
%

The derivative of the potential exhibits a maximum at the inflection point,
\begin{equation}
{\chi_{\rm max} \over f}= - \arccos \left (  {    2-p \over p   }\right).
\end{equation}
Since $m_Q\propto (U')^{1/3}$, the tensor power spectrum exhibits a peak at that time.
In this section, therefore, our reference mode $k_*$  exits the horizon at time
such that $\chi(t_*) = \chi_{\rm max}$. 
Taylor expanding around this point we get
\begin{equation}
\chi(t)\simeq\chi(t_*) + \dot\chi(t_*) (t-t_*) =  f \left [ -\arccos \left (  {    2-p \over p   }\right) +{ 2   \xi_* \over \lambda}H(t-t_*) \right ]
\end{equation}
We can now re-write $m_Q$ to lowest order in $(t-t_*)^2$ as
\begin{equation}
m_Q(t) =m_* \left [ 1-  \left ({H (t-t_*)\over \Delta N} \right )^2 \right ]
\end{equation}
where
\begin{equation}
m_*  =  \left ( \frac{g^2 \mu ^4 2^{\frac{p}{2}-1} \left(\frac{p-1}{p}\right)^{\frac{p-1}{2}} \sqrt{p}}{3 H^4 \lambda } \right )^{1/3}
\end{equation}
and 
\begin{equation}
\Delta N =   \sqrt {3   \over  p } {\lambda\over \xi_*}
\end{equation}
We now use $H(t-t_*) = \ln (k/k_*)$ and the effective mass parameter becomes
\begin{equation}
m_Q^{\rm gecos} \simeq m_*\left (1- {\ln^2(k/k_*)\over \Delta N^2 } \right )
\end{equation}
We see that --as one would have expected-- the results are very similar to the ones obtained using the simple cosine potential from the original (chromo-)natural inflation model.

The tensor power spectrum is also similar to the one derived in Ref.~\cite{Thorne:2017jft},
\begin{equation}
{\cal P}_h^{(s)}
  \simeq
 A_h  \exp\left[- {\ln^2\left( {k/ k_*}\right ) \over 2\sigma_h^2} \right]
 \label{eq:Ph_generalcosine}
\end{equation}
with
\begin{equation}
 A_h  \equiv {H^4 m_*^4 \over \pi^2 g^2 M_{\rm Pl}^4}   \mathcal{F}^2(m_*)
 \, , \quad 
 \sigma_h^2 = {\Delta N^2 \over 4 ( \alpha m_* +2)}
  \label{eq:Ah_sigmah_generalcosine}
\end{equation}
The gaussian form of the power spectrum for gravitational waves shown in Eq.~\eqref{eq:Ph_generalcosine} appears to be a general prediction of a spectator chromo-natural sector if the axion probes  a single inflection point of its potential during inflation. This goes beyond the cosine potential considered in Ref.~\cite{Dimastrogiovanni:2016fuu}, making the gaussian tensor spectrum a universal prediction of Type II potentials.

\begin{figure}
\centering
\includegraphics[width=.45\textwidth]{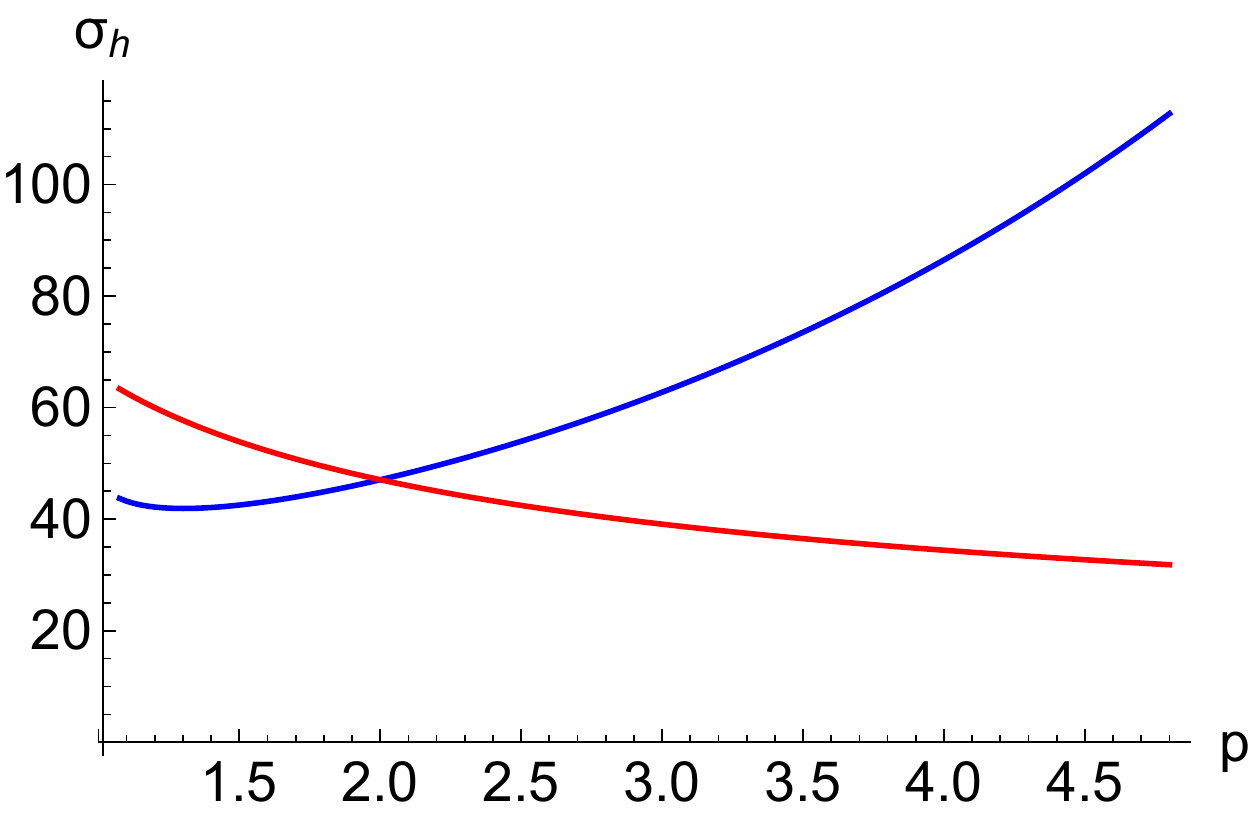}
\includegraphics[width=.45\textwidth]{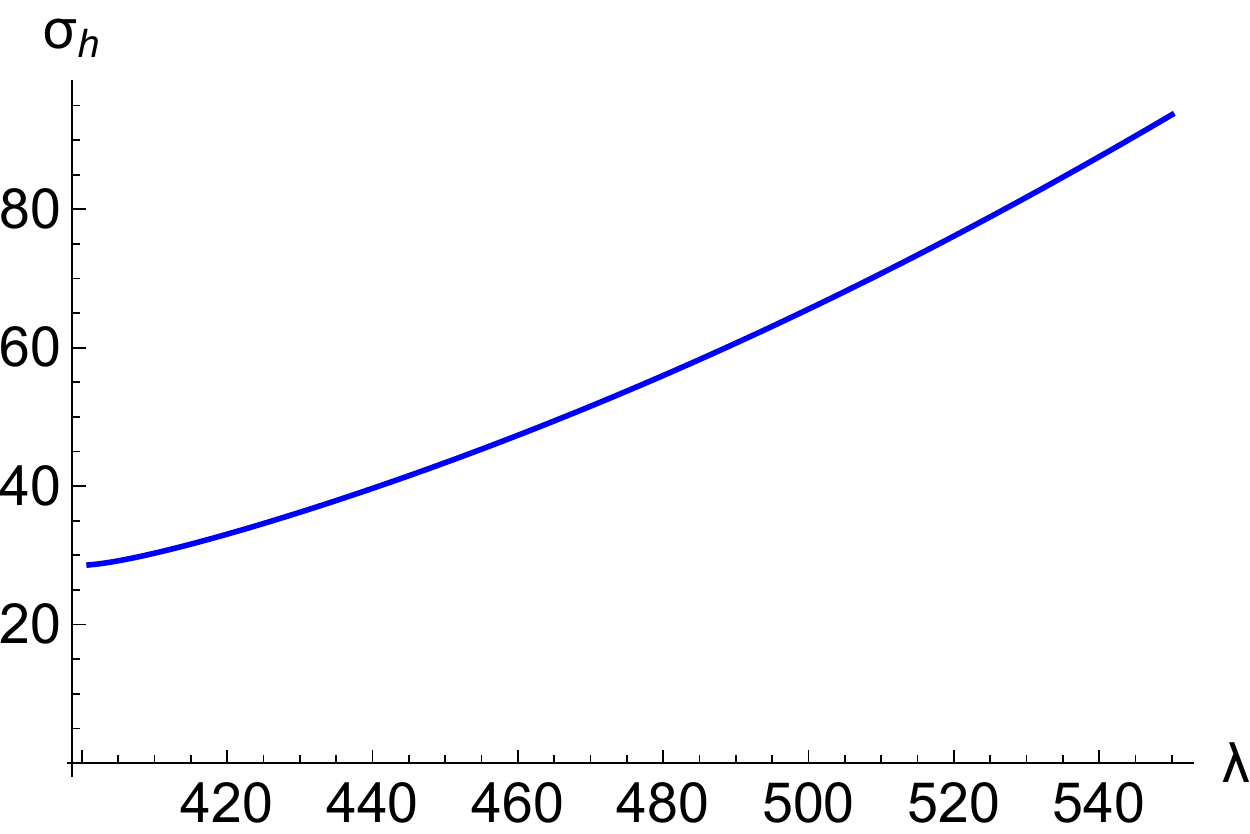}
\caption{
{\it Left:} The standard deviation $\sigma_h$ of the gaussian gravitational power spectrum for varying $p$ and correspondingly $\lambda$ (blue) or  $g$ (red), keeping all other parameters fixed. 
{\it Right:} The standard deviation $\sigma_h$ for varying $\lambda$ and  $g$, keeping all other parameters fixed and $p=2$. }
\label{fig:stdevVSgVSlambda}
\end{figure}

    This universal prediction makes it also difficult to infer the exact form of the axion potential from an observation of the power spectrum alone. As shown in Eq.~\eqref{eq:Ah_sigmah_generalcosine}, the resulting power spectrum is characterized by three parameters $A_h, \sigma_h$ and $k_*$, while the underlying models has more $H, p, \lambda, g$ and $\chi_*/f$. This leads to an obvious degeneracy between different model parameters, that produce the same gravitational wave spectrum, at least to lowest order in slow roll.
Fig.~\ref{fig:stdevVSgVSlambda}  shows the variation of $\sigma_h$ by  changing $p$ (concurrently changing $\lambda$ or $g$ to keep the amplitude $P_{h,{\rm max}}^{(s)}$ constant) or by keeping $p=2$ and changing $\lambda$ and $g$.
It would be interesting to examine if the tensor Bispectrum  \cite{Agrawal:2017awz} is able break this degeneracy and allow for a better reconstruction of Type II axion potentials based on future GW observations.

Before we conclude, we must note that  the slow-roll approximation becomes less accurate for a smaller $\Delta N$.
Ref.~\cite{Thorne:2017jft} addressed this issue for $p=2$ and $m_*=4$, and found that the agreement between numerical and analytic results is excellent for $\Delta N=10$, while the relative error becomes more than a few percent for $\Delta N=5$. We expect the slow-roll approximation to break down for $\Delta N\lesssim 1$, signaling a very sharply peaked gaussian power spectrum.



\subsection{Polynomial -- trigonometric potentials}
\label{sec:trigpotentials}

The generalized cosine potential given in Eq.~\eqref{eq:Vrobert} points towards a universal  prediction (and corresponding degeneracy) for the gaussian form of the tensor spectrum in models where the spectator axion field crosses its inflection point ($|U'|={\rm max.}$) around the time when the observable modes exit the horizon. It is interesting to connect the phenomenology of the generalized cosine potential to more complicated axion constructions. As an example, we will use the potentials derived in Ref.~\cite{Kobayashi:2015aaa}  in the context of Type IIB superstring theory compactified on the Calabi-Yau manifold
\begin{equation}
\label{eq:VtypeIIB}
\begin{split}
V_1 &= \mu^4   \left [
1-\cos\left ({ \chi\over f}\right) + \alpha\, \frac{\chi}{f} \sin \left ({ \chi\over f}\right) 
\right ]  
\, ,
\\
V_2 &= \mu^4  \left (
\frac{\chi^2}{f^2} + \alpha\, \frac{\chi}{f} \sin\left ({\chi\over f}\right )
+ \beta  \left [  1 - \cos \left ( { \chi \over f } \right )\right ]
\right )
\, ,
\end{split}
\end{equation} 
where the various terms in each potential $V_{1,2}$ are considered to be comparable. We note that the potentials $V_{1,2}$ are symmetric with respect to the minimum at $\chi=0$.

\begin{figure}
\centering
\includegraphics[width=.45\textwidth]{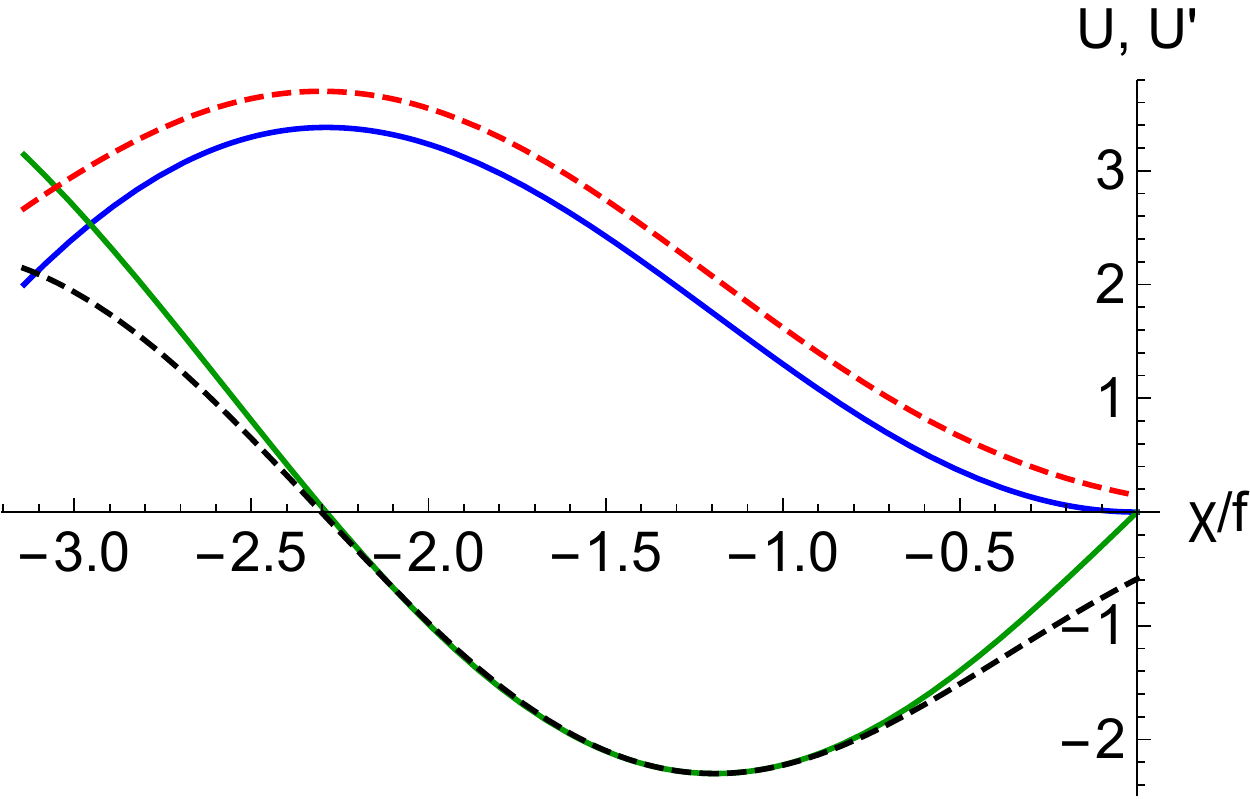}
\includegraphics[width=.45\textwidth]{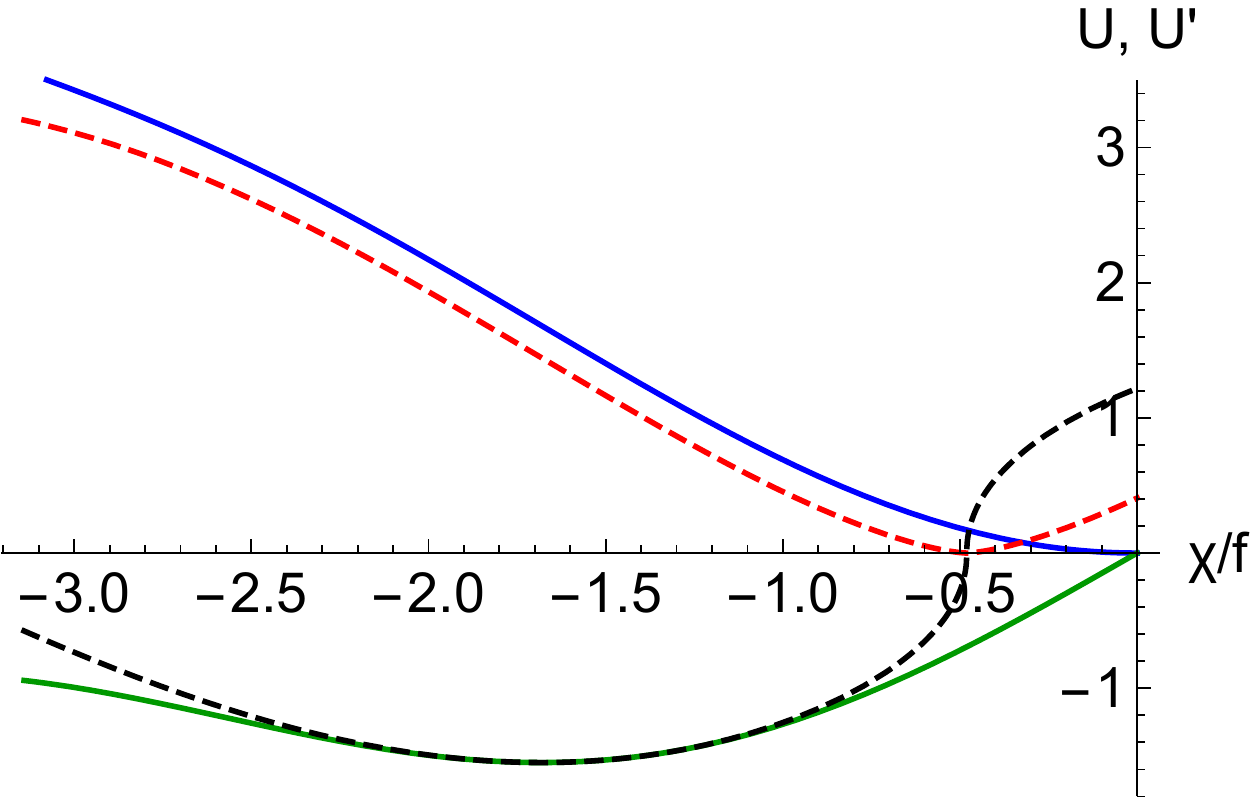}
\caption{Comparison of $U$ (blue, red) and $U'$ (green, black) for the generalized cosine potential (dashed) and the string-inspired potentials (solid) of Eq.~\eqref{eq:VtypeIIB}.
{\it Left:} The potential $V_1$ with $\alpha=1$ and the generalized cosine potential with $p\simeq 3.6$.
{\it Right:} The potential $V_2$ with $\alpha=\beta=1$ and the generalized cosine potential with $p\simeq1.47$. We see that the first derivative of the two potentials is identical for properly chosen parameters in the vicinity of the inflection point of $U$.
 }
\label{fig:compareStringV}
\end{figure}

Fig.~\ref{fig:compareStringV} shows the comparison between the potentials of Eqs.~\eqref{eq:Vrobert} and \eqref{eq:VtypeIIB} for a specific parameter choice. We see that the first derivatives of the two potentials behave similarly near the inflection point, hence the lowest-order Taylor expansion used in our analysis will be unable to distinguish between the two.

We perform numerical calculations, following the details of Appendix A, in which we keep the Hubble scale fixed, ignoring the specific form of the inflaton potential. We chose the parameters of the axion-gauge sector as follows.
\begin{equation} 
\label{parameter set 2}
\lambda=300,\qquad\quad
\mu = 1.4\times10^{15}\GeV,\qquad\quad
\left |\chi_{\rm in}\right |=1.55\times f,
\end{equation}
for the case of potential $V_1$ with $\alpha=1$, as in Fig.~\ref{fig:compareStringV}
and
\begin{equation} 
\label{parameter set 3}
\lambda=300,\qquad\quad
\mu = 1.1\times10^{15}\GeV,\qquad\quad
\left |\chi_{\rm in}\right |=2\times f,
\end{equation}
for the case of potential $V_2$ with $\alpha=\beta=1$, as in Fig.~\ref{fig:compareStringV},
where the rest of the parameters, namely $g, f, H$, are the same as Eq.~\eqref{parameter set 1}. 
The most important difference from Eq.~\eqref{parameter set 1} is the change in the Chern-Simons coupling constant $\lambda$, which is chosen to be lower for Type II potentials compared to the Type I case. Reducing the Chern-Simons coupling, reduces the extra friction term of the axion field due to the presence of the gauge sector, and thus the axion rolls down its potential faster for smaller values of $\lambda$. By performing the calculation using the parameters of Eq.~\eqref{parameter set 1}, the axion fields rolls slowly enough, that it only probes the region of the potential, where the axion potentials $V_{1,2}$ perfectly match to the corresponding generalized cosine potential. Hence the resulting gravitational wave spectra are indistinguishable. Thus we reduced the value of $\lambda$ in order for the axion field to probe a large enough distance in field space, including the region $\chi \gtrsim -0.5 f$, where the two potentials differ significantly, as shown in Fig.~\ref{fig:compareStringV}.

\begin{figure}
\centering
\includegraphics[width=.48\textwidth]{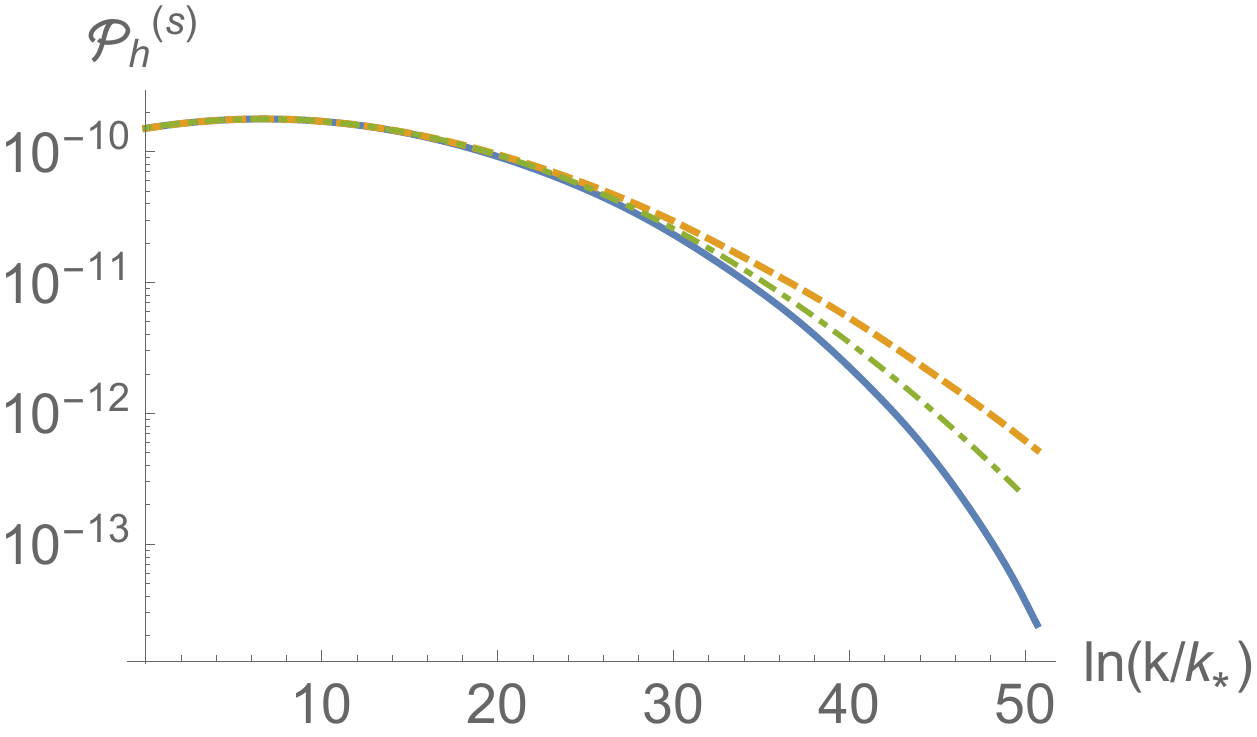}
\includegraphics[width=.48\textwidth]{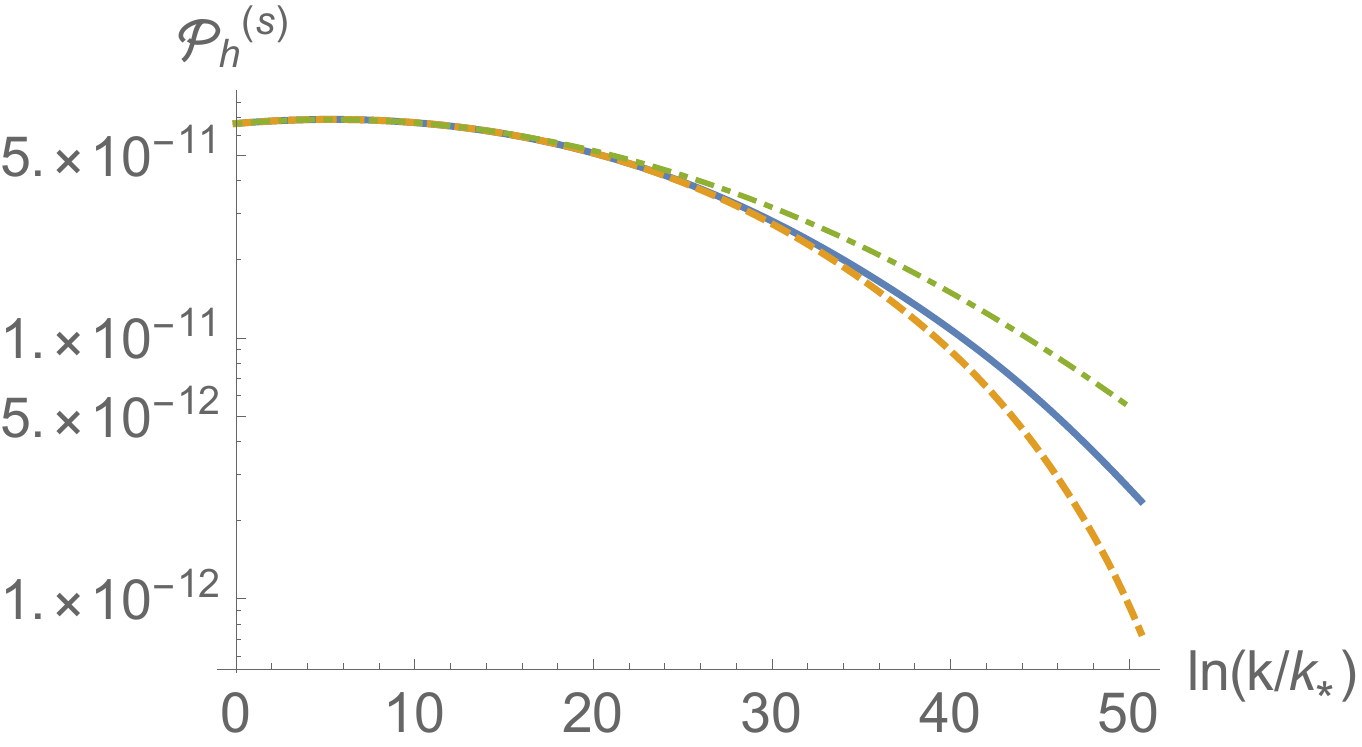}
\caption{
Comparison of the gravitational power spectrum for the 
 generalized cosine potential (orange dashed) of Eq.~\eqref{eq:Vrobert} and the string-inspired potentials (solid blue) of Eq.~\eqref{eq:VtypeIIB}.
{\it Left:} Tensor spectrum for the potential $V_1$ with $\alpha=1$ and the generalized cosine potential with $p\simeq 3.6$.
{\it Right:} Tensor spectrum  for the potential $V_2$ with $\alpha=\beta=1$ and the generalized cosine potential with $p\simeq1.47$. 
The green dot-dashed line shows a Gaussian fit near the maximum of the tensor spectrum.
 }
\label{fig:compareStringVspectrum}
\end{figure}

The resulting gravitational wave power spectra are shown in Fig.~\ref{fig:compareStringVspectrum} and are worth closer examination. We see that the computed spectra derived by using either $V_{1,2}$ or the generalized cosine potential match extremely well near the maximum. Away from the maximum the generalized cosine potential either overestimates (in case of the specific realization of $V_1$) or underestimates (in case of the specific realization of $V_2$) the power spectrum arising from the sting-inspired potentials of Eq.~\eqref{eq:VtypeIIB}. This is easy to understand by looking at Fig.~\ref{fig:compareStringV}, where the potential derivative $dU/d\chi$ of the generalized cosine potential is below that of $V_1$ and above that of $V_2$ for low values of $|\chi/f|$, occurring later into inflation and thus corresponding to smaller scales (larger values of $k/k_*$).

An interesting point is the possibility of a spectator axion-gauge sector to produce a gravitational wave spectrum peaked at an arbitrary scale. For a cosine-type potential, or any potential with a single inflection point, the axion will generally transverse a distance $\Delta \chi/f = {\cal O}(1)$ during inflation. If inflation last much more than $60-70$ $e$-folds, the axion field would have most likely settled to its minimum, before the observable scales exited the horizon during inflation. In order to produce a sharply peaked tensor spectrum at small scales, for example within the LISA band, the axion would have to cross its inflection point at a very specific time, which would require significant fine-tuning, for example by setting the axion initially very close to the maximum of its potential. One way to overcome this fine-tuning is by invoking some waterfall-type transition for the axion, introducing an inflaton-dependent mass term, like the one found in models of hybrid inflation \cite{Linde:1993cn, Guth:2012we, Halpern:2014mca, Lyth:2012yp, Clesse:2015wea}. This makes the model inherently more complicated, by directly coupling the inflaton and axion sectors in a very specific way. Furthermore, if the peak of the tensor power spectrum is produced at small scales, the axion might not have time to relax to its minimum before the end of inflation, leading to the curvaton-like situation that was discussed in Section~\ref{sec:powerlaw}. The viability of Type II models with tensor spectra peaked late into inflation will ultimately depend on the details of the reheating epoch and the time evolution of the energy density in the inflaton and axion sectors after inflation.

Before we conclude this section, we must note that  the potentials of Eq.~\eqref{eq:VtypeIIB} can describe all three types of axion potentials that we study: power-law, single inflection point and axion monodromy, depending on the magnitude of the dimensionful parameters $\alpha$ and $\beta$.  For example $V_2$ for $\alpha,\beta\to 0$ is a simple  power law (quadratic potential), whereas if either $\alpha $ or $\beta$ is subdominant to the power-law term, but not negligible, we recover a form of modulated potential, falling within the Type III potentials of Table~\ref{tab:Types}. Type III potentials  are studied in detail in Section~\ref{sec:monodromy}.


\section{Monodromy Potential}
\label{sec:monodromy}

In this section, we consider the monodromy potential, denoted as Type III in Table~\ref{tab:Types},
\begin{equation}
U(\chi)=\mu^4\left[\,\left|\frac{\chi}{f}\right|^p+
\delta \cos \left(\frac{\kappa \chi}{\delta f}\right)
\right],
\label{eq:monodromydef}
\end{equation}
where $\delta$ and $\kappa$ are two dimensionless parameters encoding the amplitude and frequency of the oscillatory part of the potential.
In this case, we cannot simply apply the formulae of Section~\ref{Slow Roll background}, because the oscillatory motion invalidates the slow-roll approximation.%
\footnote{In Ref.~\cite{Obata:2016xcr}, Chromo-Natural inflation with an oscillating potential was studied. The sourced GW power spectrum and its detectability by interferometers were discussed based on the slow-roll approximation
which is no longer valid for a oscillatory potential, unless the oscillation is sufficiently slow. }
However, we can still analytically solve the background dynamics,
if the oscillatory components in the potential and its derivative are subdominant
(i.e. $\delta$ and $\kappa$ are small).
Hereafter, for simplicity and definiteness, we consider the case with $p=1$, namely the linear potential case. The generalization to $p\neq 1$ is expected to be straightforward, while more complicated expressions for the resulting gravitational wave template may be obtained.

\subsection{Slow-roll background revisited}
\label{sec:TypeIIISR}

We decompose the background fields $\chi(t)$ and $Q(t)$ into two parts,
\begin{equation}
\chi=\chi_0+\chi_\osc \, ,
\qquad
Q=Q_0+ Q_\osc \, , 
\end{equation}
where $\chi_0$ and $Q_0$ satisfy the slow-roll equations and all the results of Section~\ref{Slow Roll background} are applicable.
The terms $\chi_\osc$ and $Q_\osc$ are treated as perturbations around $\chi_0$ and $Q_0$ caused by the oscillating feature in the potential.
Thus the potential is also decomposed into two parts:
\begin{equation}
U_0\equiv- \mu^4 \frac{\chi}{f},
\qquad
U_\osc\equiv \mu^4 \delta \cos \left(\frac{\kappa \chi}{\delta f}\right).
\end{equation}
Note a minus sign is introduced to $U_0$ under the assumption of a negative initial value for $\chi$, $\chi_{\rm in}<0$.
The non-oscillatory part satisfies
\begin{equation}
U'_0
\simeq-\frac{3g\lambda}{f}HQ^3_0,
\qquad
2H^2 Q_0
+2g^2 Q^3_0 \simeq\frac{g\lambda}{f} Q^2_0 \dot{\chi}_0 \, ,
\end{equation}
where $m_Q$ and $\xi$ are constant at leading order:
\begin{equation}
m_0= \left(\frac{g^2\mu^4}{3\lambda H^4}\right)^{\frac{1}{3}},
\qquad 
\xi_0=m_0+m_0^{-1}.
\label{m0 xi0}
\end{equation}

Now we perturbatively solve the EoMs for $\chi_\osc$ and $Q_\osc$, Eqs.~\eqref{Full EoM chi} and \eqref{Full EoM Q}. 
\begin{itemize}
\item We begin with Eq.~\eqref{Full EoM chi}. Even for the modulated potential of Eq.~\eqref{eq:monodromydef}, $\ddot{\chi}_\osc$ and $3H\dot{\chi}_\osc$ are negligible in the equation of motion (as we check later around Eq.~\eqref{ddot negligible}), and the linearized equation is written as
\begin{equation}
U'_\osc(\chi_0)
= -\frac{9g\lambda}{f}HQ^2_0 \left(Q_\osc+\frac{\dot{Q}_\osc}{3H}\right),
\label{Uosc EoM}
\end{equation}
where we also ignore the term proportional to $\dot{Q}_0/HQ_0\ll1$.
Note that the argument of the $U'_\osc$ in the left hand side is approximated by $\chi_0$. This equation can be solved as
\begin{align}
Q_\osc&=\kappa a^{-3}\int \dd t \, a^3HQ_0 \sin\left(\frac{\kappa \chi_0}{\delta f}\right),
\notag\\
&\simeq \frac{\kappa}{3}Q_0
\left[1+\left(\frac{\kappa \dot{\chi}_0}{3H\delta f}\right)^2\right]^{-1}
\Bigg( \sin\left(\frac{\kappa \chi_0}{\delta f}\right)
- \frac{\kappa \dot{\chi}_0}{3H\delta f} \cos\left(\frac{\kappa \chi_0}{\delta f}\right)\Bigg),
\end{align}
where the time variation of $H, Q_0$ and $\dot{\chi}_0$ is neglected.
Introducing two new dimensionless parameters,
\begin{equation}
\omega\equiv \frac{\kappa \dot{\chi}_0}{H \delta f}=\frac{2\kappa}{\delta \lambda}\xi_0,
\qquad
\Delta \equiv \frac{\kappa }{\sqrt{9+\omega^2}},
\end{equation}
the above result can be rewritten as
\begin{equation}
\frac{Q_\osc}{Q_0} \simeq \Delta \sin(\omega H t),
\label{Q osc}
\end{equation}
where we dropped a constant phase of the sinusoidal function.

\item We next solve the EoM for $Q$, Eq.~\eqref{Full EoM Q}.
Linearizing it in terms of $Q_\osc$ and $\chi_\osc$, we obtain
\begin{equation}
\ddot{Q}_\osc+3H\dot{Q}_\osc +2H^2 Q_\osc 
+6g^2 Q_0^2 Q_\osc=\frac{g\lambda}{f} \left(2Q_0 \dot{\chi}_0 Q_\osc+Q_0^2\dot{\chi}_\osc\right),
\end{equation}
where we have used the slow-roll condition $\dot{H}\ll H^2$. This equation is easily solved with respect to $\dot{\chi}_\osc$. Using Eq.~\eqref{Q osc},
we obtain 
\begin{equation}
\xi_\osc \equiv \frac{\lambda \dot{\chi}_\osc}{2f H}=
\frac{\Delta}{2m_0}\Big[(2m_0^2-\omega^2-2)\sin(\omega H t)+3\omega \cos(\omega H t)\Big],
\end{equation}
where we have used $\xi_0=m_0+m_0^{-1}$.
%
%
\end{itemize}
In summary, we have derived the following analytic expressions,
\begin{align}
m_Q&=m_0+m_\osc=m_0\left[1+\Delta \sin(\omega H t)\right],
\label{Monodromy mQ}
\\
\xi &=\xi_0+\xi_\osc=m_0
+ m_0^{-1}\left[1+
\frac{\Delta}{2}\Big\{(2m_0^2-\omega^2-2)\sin(\omega H t)+3\omega \cos(\omega H t)\Big\}\right].
\end{align}
The slow-roll relation $\xi= m_Q+m_Q^{-1}$ that holds for non-oscillatory potentials, 
is not exact in the case of axion monodromy,
%
\begin{equation}
 \frac{\xi}{m_Q+m_Q^{-1}} =1+\frac{\omega\sqrt{9+\omega^2}}{2(m_0^2+1)}\Delta \sin(3\omega H t+\theta)
+\mathcal{O}(\Delta^2)\, ,
\label{xi m ratio}
\end{equation}
where $\theta$ is a constant phase. Thus if $\omega\Delta\ll m_0^2$, the slow-roll relation is approximately satisfied.
In a similar way, the time evolution of $Q(t)$ is evaluated as
\begin{equation}
\frac{\dot{Q}}{H Q} \simeq \frac{\dot{Q}_\osc}{H Q_0}\simeq \omega \Delta \cos(\omega H t).
\label{epsilon Q}
\end{equation}
Therefore, if $\omega\Delta\ll1$, the time variation of $Q$ and $m_Q$ is small.

\begin{figure}[tbp]
    \hspace{-2mm}
  \includegraphics[width=70mm]{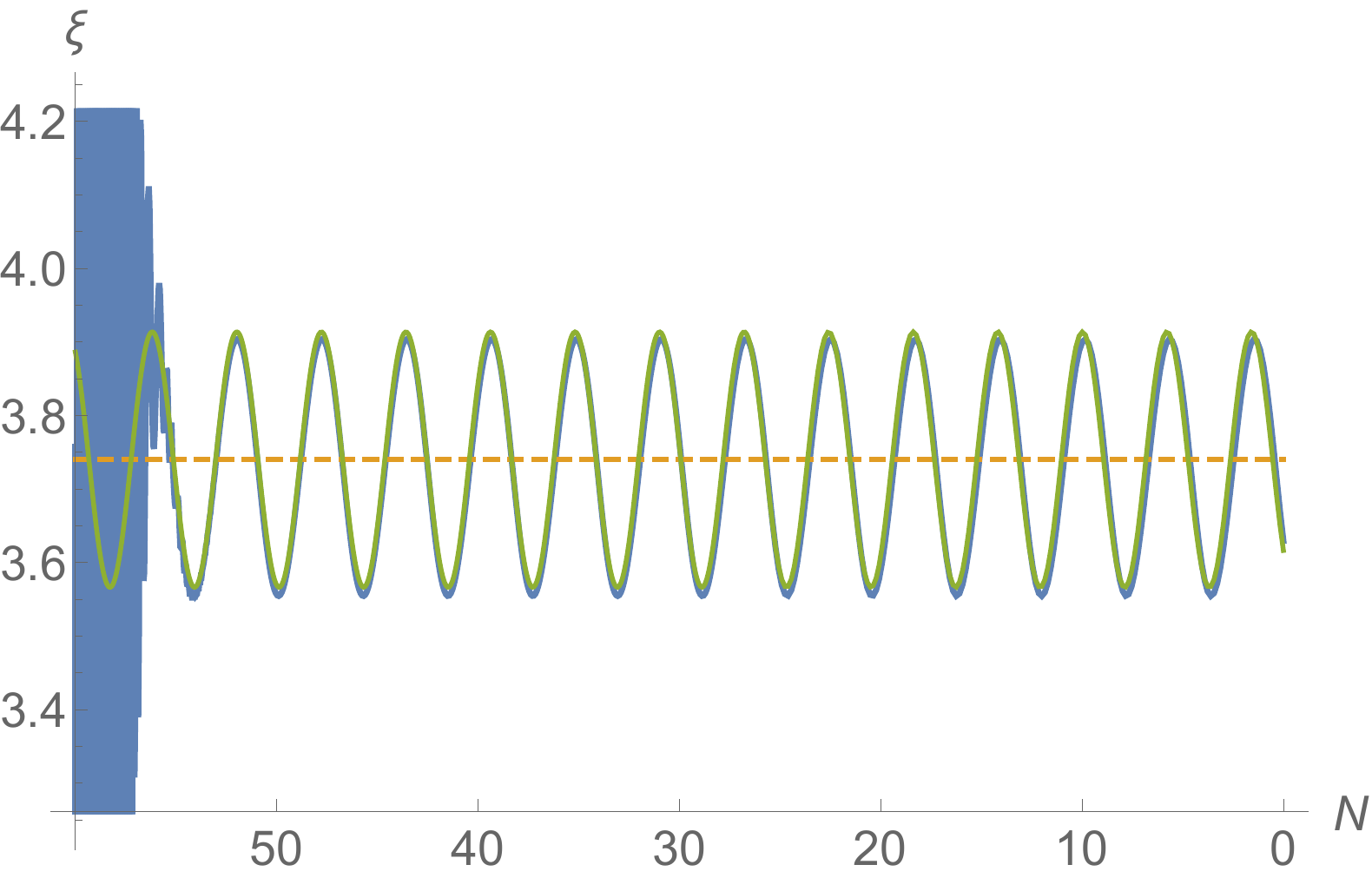}
  \hspace{5mm}
  \includegraphics[width=70mm]{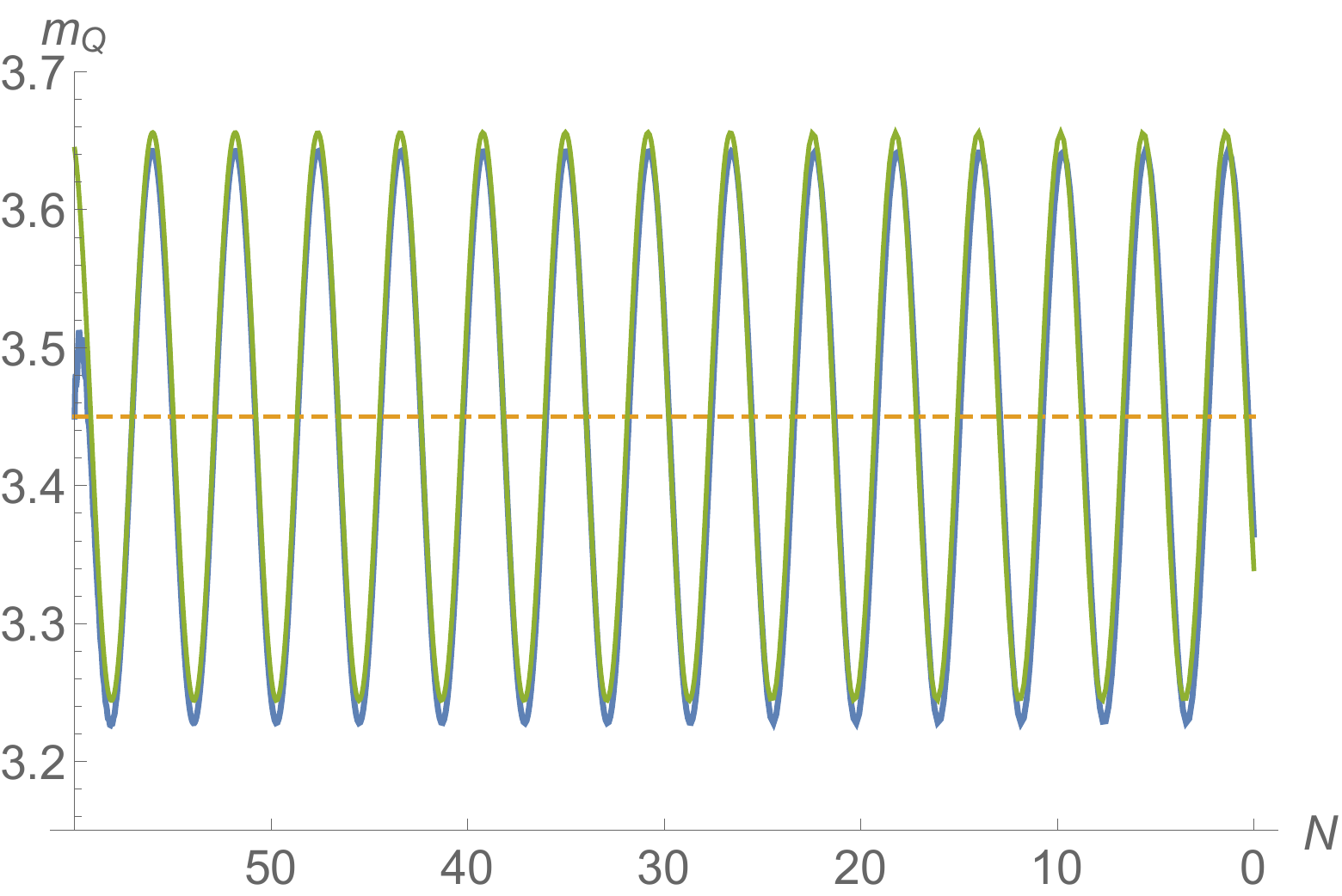}
  \caption
 {{\bf(Left panel)} The numerical result for $\xi$ (blue line) and the analytic expression $\xi=\xi_0+\xi_\osc$ (green line) are compared. The leading order non-oscillatory term $\xi_0$ is shown as a yellow dashed line. {\bf(Left panel)} The numerical result for $m_Q$ (blue line) and the analytic expression $m_Q=g(Q_0+Q_\osc)/H$ (green line) are compared. The leading order  term $m_0$ is shown as a yellow dashed line.
 The  accuracy of the analytical expressions is evident.
 }
 \label{xi mQ plot}
\end{figure}

To verify the above analytic derivation, we perform numerical calculations using the parameters of Eq.~\eqref{parameter set 1} with the exception of the initial axion amplitude
$\chi_{\rm in}=-5.7 \times 10^{16}\GeV$ and the new parameters
\begin{equation}
\delta=\frac{1}{500},\qquad \kappa=\frac{1}{5} \, .
\label{dk parameters}
\end{equation}
With these parameters, we have 
\begin{equation}
m_Q\approx 3.45\, ,
\quad \xi_0\approx 3.74\, ,
\quad \omega \approx1.5\, ,
\quad \Delta \approx 0.06\, .
\label{oscillatory parameters}
\end{equation}
The oscillating period is given by $2\pi/\omega \approx 4.2$ in units of e-folding number.
In Fig.~\ref{xi mQ plot} we compare
the analytically obtained formulae and the numerical results.%
\footnote{The phases of the oscillation in $\chi_\osc$ and $Q_\osc$ are not fixed by our analytic calculation. In Fig.~\ref{xi mQ plot}, therefore, we have introduced a common constant phase of the trigonometric functions and determined it such that the analytical and numerical results oscillate in phase around $N\simeq 50$.}
An excellent agreement can be seen.
Since we take into account only the first order corrections to $\chi_\osc$ and $Q_\osc$ with respect to $\Delta$,
we expect errors are $\mathcal{O}(\Delta^2) \approx 0.4\%$.
Indeed, the relative errors of the analytic expressions are sub-percent
for more than $40$ e-folds in the both panels in Fig.~\ref{xi mQ plot}. 

%
%

Before concluding this section, we must check the consistency of ignoring $\ddot{\chi}_\osc$ and $3H\dot{\chi}_\osc$  in Eq.~\eqref{Uosc EoM}.
One can show that
\begin{equation}
\frac{3H\dot{\chi}_\osc}{U'_\osc(\chi_0)}\simeq \left ({1\over \Lambda_0} \right)^{2}\,
\frac{(2m_0^2-\omega^2+2)\sin(\omega H t)+3\omega \cos(\omega H t)}{m_0^2\sqrt{9+\omega^2}\sin(\omega H t)}.
\label{ddot negligible}
\end{equation}
Since $\Lambda_0\equiv \lambda Q_0/f$ is assumed to be very large (see Eq.~\eqref{Lambda def}), $3H\dot{\chi}_\osc$ is confirmed to be negligible. As a concrete example we must note the value $\Lambda_0\approx 50$, which is derived from the parameters of Eq.~\eqref{parameter set 1} that we used.
The ratio of $\ddot{\chi}_\osc/U'_\osc$ shows a similar parameter dependence, with  an extra multiplicative factor of $\mathcal{O}(\omega)$, which does not change the order of magnitude. Therefore, we are justified to ignore $\ddot{\chi}_\osc$ and $3H\dot{\chi}_\osc$, since they represent a subdominant contribution to the equation of motion compared to $ U'_{\rm osc}$.

\subsection{Oscillatory tensor power spectrum}

\begin{figure}[tbp]
    \hspace{-2mm}
  \includegraphics[width=70mm]{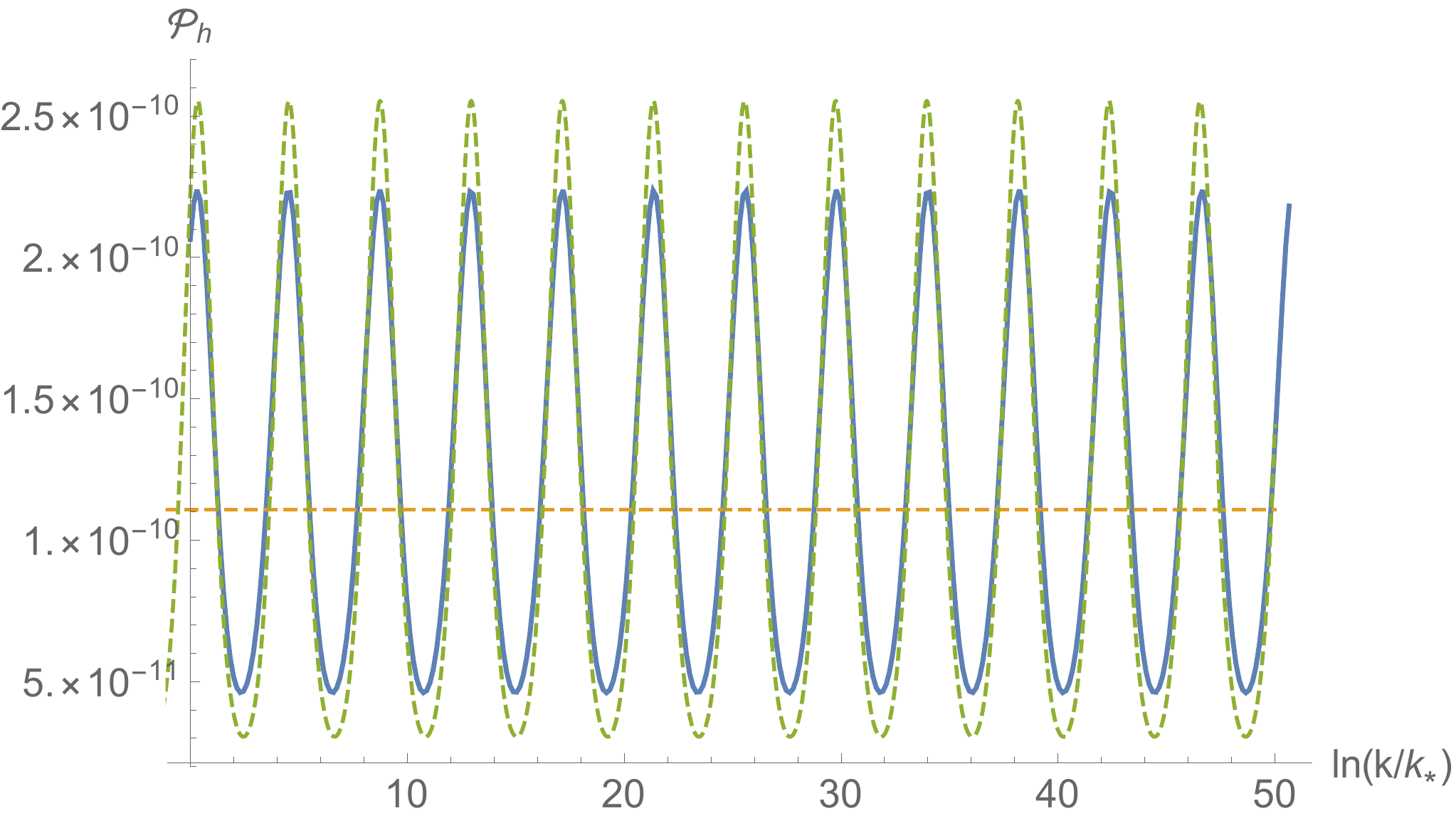}
  \hspace{5mm}
  \includegraphics[width=70mm]{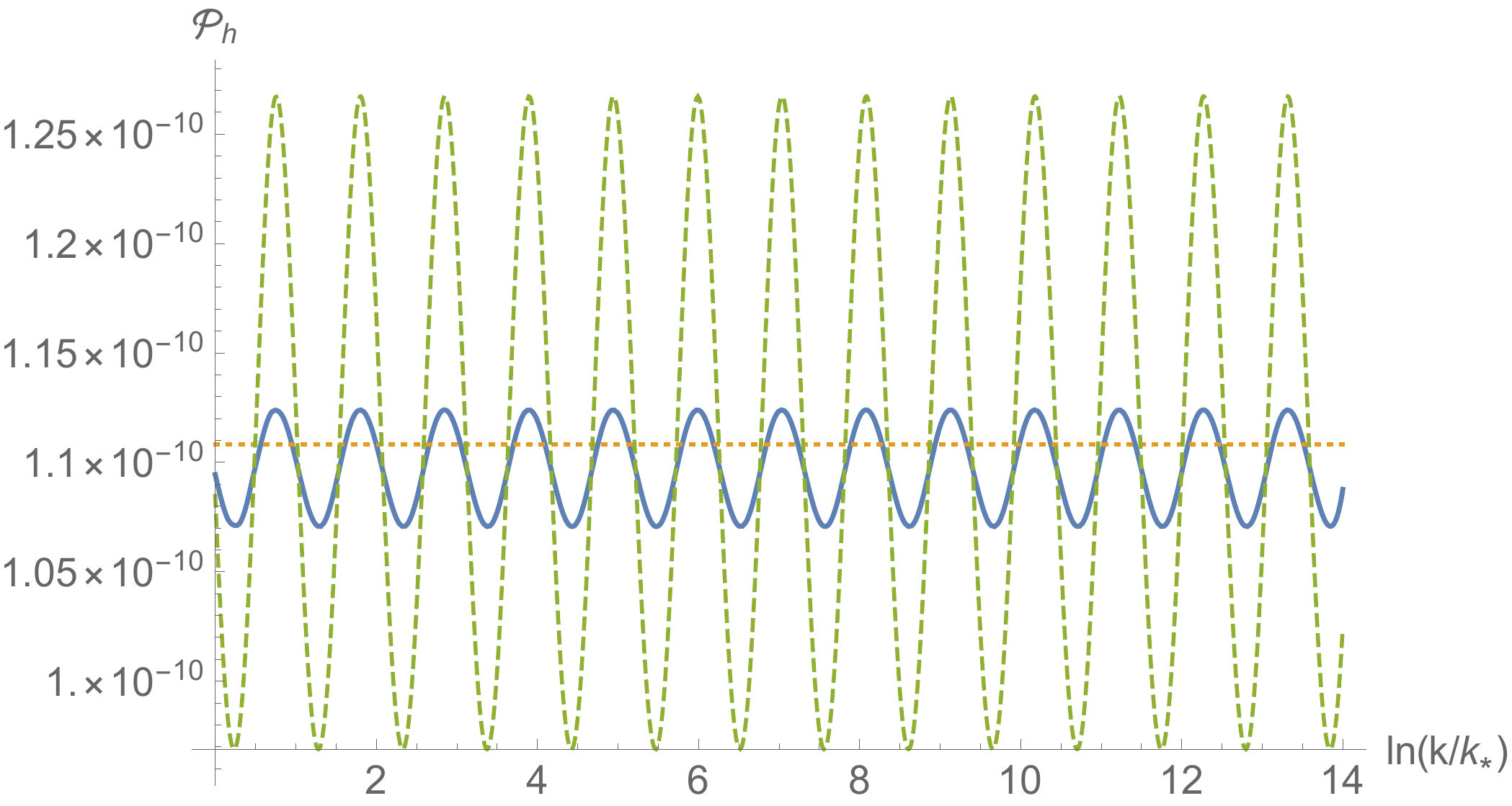}
  \caption
 {We compare the numerical result (blue line) and the analytic expression Eq.~\eqref{analytic Ph} (green dashed line). The parameters are the same as Eq.~\eqref{parameter set 1}, and we set
$\omega \approx1.5, \Delta \approx 0.06$ (left panel) and $\omega \approx 6, \Delta \approx 7.5\times 10^{-3}$  (right panel). Note that the oscillation is much faster in the right panel and the relative discrepancy is also larger, although both the oscillation amplitude and the time variation of $m_Q$ which is characterized by $\omega\Delta$  are smaller in the right panel. The yellow dashed lines represent the analytic result without the oscillation, $\mcP_h=H^4m_0^4 \exp\left[2\alpha m_0\right]/(\pi^2 g^2 \Mpl^4)$,
whose deviation from the numerical mean value is small.
}
 \label{Mon Phs}
\end{figure}

In the case of the axion monodromy potential with an oscillatory part, the slow-roll approximations are partially violated as we saw in  Section~\ref{sec:TypeIIISR}, meaning that the conditions $\dot{Q}\ll HQ$ and $\xi\simeq m_Q+m_Q^{-1}$ are modified. Although, in general, the expression for $\mcP_h^{(s)}$ given in Eq.~\eqref{Phs slowroll}
should be re-derived without the slow-roll approximation, we can still exploit it if the oscillatory part is subdominant. 
Plugging $m_Q=m_0[1+\Delta \sin(\omega H t)]$ into Eq.~\eqref{Phm}, we obtain
\begin{align}
\mcP_h^{(s)} &\simeq
\frac{H^4m_0^4}{\pi^2 g^2 \Mpl^4}\Big(1+\Delta \sin(\omega H t)\Big)^4 \exp\left[2\alpha m_0\Big(1+\Delta \sin(\omega H t)\Big)\right]
\label{analytic Ph}
\\
&\xrightarrow{\Delta\ll1}\ 
\frac{H^4m_0^4}{\pi^2 g^2 \Mpl^4}\Big(1+4\Delta \sin(\omega H t)\Big) \exp\left[2\alpha m_0\Big(1+\Delta \sin(\omega H t)\Big)\right]
\label{half approx Ph}
\\
&\xrightarrow{2\alpha m_0 \Delta\ll1}\ 
\frac{H^4m_0^4}{\pi^2 g^2 \Mpl^4}\Big(1+2(\alpha m_0+2)\Delta \sin(\omega H t)\Big)e^{2\alpha m_0}.
\label{approx Ph}
\end{align}
Thus if $\alpha m_0\Delta \ll 1$, the GW power spectrum oscillates solely due to the sinusoidal term in the overall prefactor in Eq.~\eqref{approx Ph}. On the other hand, if $\Delta$ is not sufficiently small, such that $\alpha m_0 \Delta \gtrsim1$, the exponential function in Eq.~\eqref{half approx Ph} cannot be expanded and the oscillation of $\mcP_h$ takes a distinct form in which sinusoidal functions appear not only in the prefactor but also in the exponent.\footnote{In the left panel of Fig.~\ref{Mon Phs}, this condition is not well satisfied $2\alpha m_0 \Delta\approx 0.84$ and a deviation from sinusoidal oscillation can be seen. In the right panel, however, $2\alpha m_0 \Delta\approx 0.1$ and we do not see the deviation.} It is an interesting possibility because the degeneracy between $\Delta$ and $\alpha m_0 \Delta$ can be resolved and
we can potentially extract more information from $\mcP_h$ in that case.
Nonetheless, we do not investigate the deviation from the sinusoidal oscillation in this paper leaving it for future work.


%
%
In Fig.~\ref{Mon Phs}, we compare the numerical result of the GW power spectra (blue solid line) to the analytic expressions (green dashed line). We use the parameters of Eq.~\eqref{parameter set 1} and the oscillatory potential is characterized by $\delta=2\times 10^{-3},  \kappa=0.2$ corresponding to $\omega \approx1.5, \Delta \approx 0.06$ (left panel) which is the same as Eq.~\eqref{dk parameters} and $\delta=1.2\times 10^{-4},  \kappa=0.05$ corresponding to $\omega \approx 6, \Delta \approx 7.5\times 10^{-3}$ (right panel).  
As one can see, the right panel in which the oscillation is faster than the left panel shows a greater discrepancy between the numerical and analytic results. Since the relative time variation of $Q$ (derived in Eq.~\eqref{epsilon Q}) are $\dot{Q}/HQ\sim \omega\Delta \approx 0.09$ (left panel) and $0.045$ (right panel), our approximate analytic expression is supposed to work better for the right panel. However, the analytic expression for $\mcP_h^{\rm (s)}$, Eq.~\eqref{Phs slowroll}, is not necessarily accurate for cases with varying background fields, even if $\dot{Q}/H Q$ is small. 
Fig.~\ref{Mon Phs} implies that for Eq.~\eqref{Phs slowroll} to work accurately, another condition than $\omega \Delta\ll 1$ is involved.   

\begin{figure}[tbp]
    \hspace{-2mm}
  \includegraphics[width=70mm]{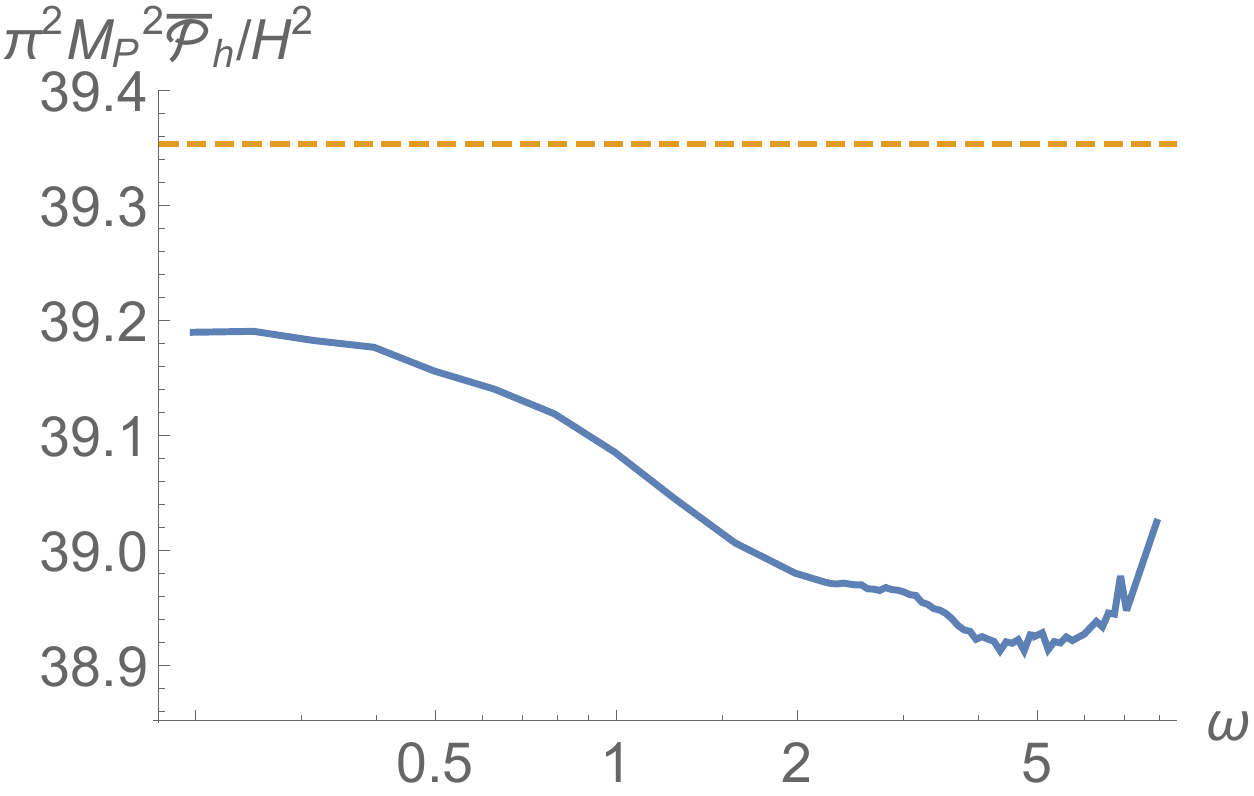}
  \hspace{5mm}
  \includegraphics[width=70mm]{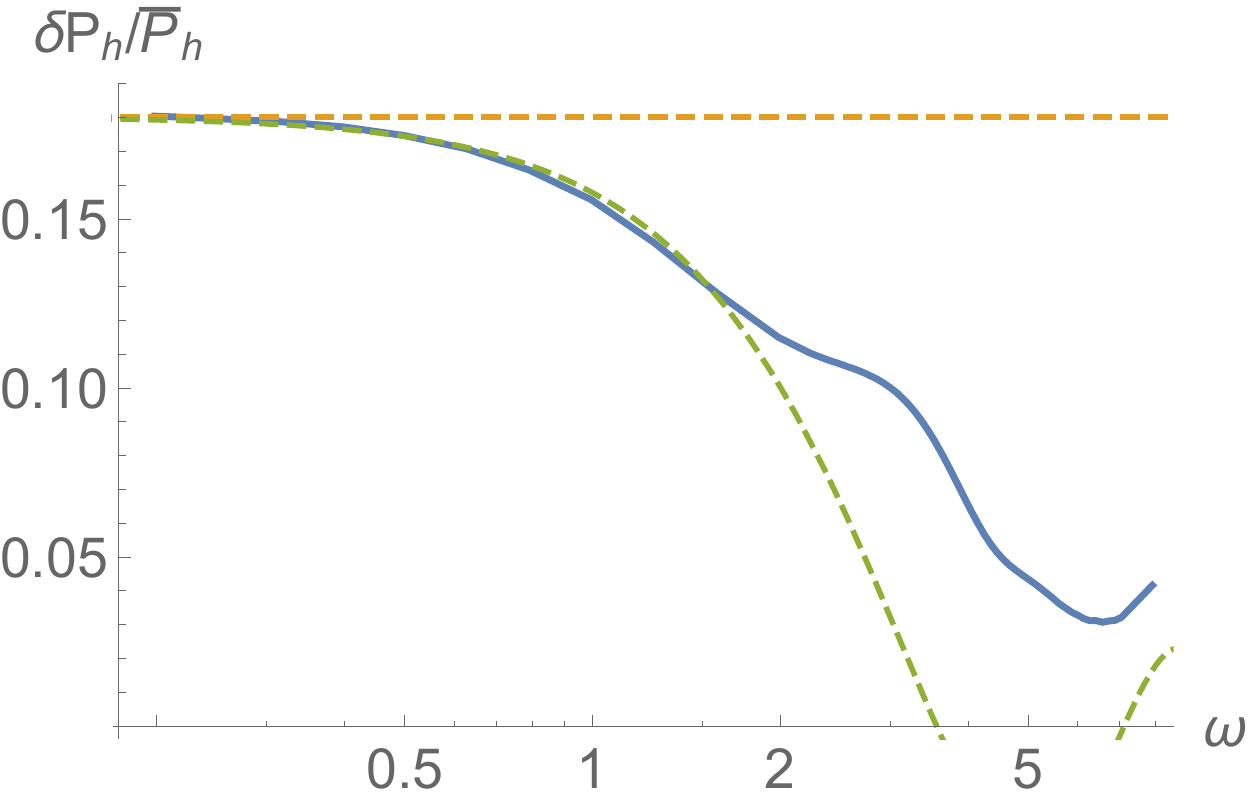}
  \caption
 {We plot how the mean value and the oscillating amplitude of the sourced GW power spectrum depend on the oscillating frequency $\omega$ in the left and right panels, respectively. In particular, the oscillation amplitude is significantly lowered for $\omega/2\pi\gtrsim 0.1$.
The yellow dashed lines denote the analytically derived expressions in Eq.~\eqref{Phs slowroll}, namely 
$\bar{\mcP}_h^{\rm (s)}= H^4 m_0^4 e^{2\alpha m_0}/(\pi^2 g^2 \Mpl^4)$
and $\delta \mcP_h/\bar{\mcP}_h= 2(\alpha m_0+2)\Delta$.
The green dashed line in the right panel includes the suppression factor $\Gamma(\omega)$ in Eq.~\eqref{S factor}, namely $\delta \mcP_h/\bar{\mcP}_h= 2(\alpha m_0+2)\Delta \times \Gamma(\omega)$.
}
 \label{Ph omega}
\end{figure}

In order to further investigate the effect of oscillatory behavior of the background fields on tensor perturbations, we perform numerical calculations for varying $\omega$ by fixing $\Delta=10^{-2}$.
In Fig.~\ref{Ph omega}, we show the dependence of the mean value (left panel) and the oscillation amplitude (right panel) of $\mcP_h^{\rm (s)}$ on the oscillation frequency $\omega$.
The mean value does not change much (less than 1\%).%
\footnote{The small ($\approx 0.4\%$) discrepancy between the numerical mean value and the analytic one is seen in Fig.~\ref{Ph omega} even in the limit $\omega\to0$. This is presumably because of  the deviation from a sinusoidal oscillation discussed below Eq.~\eqref{approx Ph} that obscures the fitting of the mean value.}
However, the oscillation amplitude significantly decreases as the frequency increases for $\omega/2\pi\gtrsim 0.1$ (see the right panel in Fig.~\ref{Ph omega}).\footnote{Note that one cannot straightforwardly extend this figure into $\omega\gg 1,$ because neat oscillations of the background fields are not realized in such parameter region and their behaviors become rather disordered.}
This is because the effective oscillation amplitudes of the background fields which the tensor perturbations effectively  feel are reduced, if the oscillation time scale gets shorter. 
As a result, Eq.~\eqref{approx Ph} does not give an accurate oscillation amplitude of $\mcP_h^{\rm (s)}$, even though the oscillation amplitudes of $m_Q$ and $\xi$ which are characterized by $\Delta$ is small enough.
%
%

Let us now consider the suppression effect of a large $\omega$ on the oscillation amplitude of $\mcP_h^{\rm (s)}$. Although it is difficult to solve the equations of motion for the tensor perturbation of the $SU(2)$ gauge field $t_{ij}$ and the gravitational waves $h_{ij}$ even if one treats their oscillatory parts as perturbation, we can derive a useful analytic expression for the suppression factor in the following simple argument. 
Examining the equation of motion for the spin-$2$ tensor degree of freedom, one can easily find that $t_{ij}(\tau,\bm k)$ becomes unstable and is amplified for %
\begin{equation}
m_Q+\xi-\sqrt{m_Q^2+\xi^2}<-k\tau<m_Q+\xi+\sqrt{m_Q^2+\xi^2},
\label{unstable time}
\end{equation}
where $\tau$ is the conformal time.
In the unit of e-folding number, this interval is given by
\begin{equation}
2\delta N=\ln\left[\frac{m_Q+\xi+\sqrt{m_Q^2+\xi^2}}{m_Q+\xi-\sqrt{m_Q^2+\xi^2}}\right]\simeq
\ln\left[5.83+\frac{2.06}{m_Q^4}+\mathcal{O}(m_Q^{-6}) \right]\simeq 1.76,
\end{equation}
where we used $\xi\simeq m_Q+m_Q^{-1}$ at the second equal and ignored the  $m_Q$ dependence at the third equal.
If the oscillation of $m_Q$ (and $\xi$) is sufficiently slow, the amplification of $t_{ij}$ can be computed by using the value of $m_Q(t)$ at anytime in Eq.~\eqref{unstable time} which is almost constant. On the other hand, if $\omega$ is not so small, the time variation of $m_Q$, namely $m_\osc(t)\propto \sin (\omega H t)$, should be taken into account. At the leading-order approximation, $m_\osc(t)$ can be replaced by the averaged value over the time interval Eq.~\eqref{unstable time}.
Since the time average of the sine function is computed as
\begin{equation}
\frac{1}{2\delta N}\int^{Ht+\delta N}_{Ht-\delta N}dN \sin (\omega N)
=\frac{\sin(\omega\delta N)}{\omega \delta N}\sin(\omega H t),
\end{equation}
we obtain a suppression factor 
\begin{equation}
\Gamma(\omega) = \frac{\sin(0.88 \omega)}{0.88\omega}.
\label{S factor}
\end{equation}
The right panel of Fig.~\ref{Ph omega} shows the analytic expression including this suppression factor as a green dashed line. 
Despite the simplicity of this argument, an excellent agreement with the numerical result can be seen for $\omega \lesssim 2.$
Consequently, if $\Delta$ is sufficiently small, $2\alpha m_0 \Delta \ll 1$, our analytic expression for the oscillatory tensor power spectrum is given by
\begin{equation}
\mcP_h^{(s)}(k) \simeq
\frac{H^4m_0^4}{\pi^2 g^2 \Mpl^4}e^{2\alpha m_0} \Big(1+
\Gamma(\omega)\,
2(\alpha m_0+2)\Delta \sin(\omega \ln(k/k_*))\Big),
\qquad
(2\alpha m_0 \Delta\ll1).
\end{equation}
%

\section{CMB Analysis}
\label{sec:CMB}

\subsection{Type I \& II potentials}

In anticipation of future experiments aimed at detecting primordial gravitational wave signals in the CMB, we must examine the possibility of determining wether the detected tensor modes correspond to vacuum fluctuations or are generated through a spectator axion-gauge sector.
It is well known that  primordial gravitational waves generated through the stretching of vacuum fluctuations during inflation obey the consistency relation, $n_T^{\rm (vac)}=-2\epsilon_H$, where $n_T^{\rm (vac)}$ is the tensor tilt and $\epsilon_H = -\dot H/H^2$ is the first slow-roll parameter. 
The tensor tilt $n_T^{\rm (vac)}$ is thus  always negative and has a typical magnitude $\mathcal{O}(10^{-2})$ or smaller.
On the contrary, the spectrum of gravitational waves that are sourced through axion-gauge dynamics or higher-order gauge interactions can have a blue tilt \cite{Adshead:2016omu, Adshead:2017hnc}. In the specific case of Type I potentials,   $n_T^{\rm power}$ is given by Eq.~\eqref{nT power} and can be positive and have an $\mathcal{O}(1)$ magnitude. 
Therefore, a large and positive spectral tilt of the tensor power spectrum can be a smoking gun for this class of models.

Nonetheless, we must note that the observed tilt can deviate from Eq.~\eqref{nT power}, if the sourced gravitational wave spectrum $\mcP_h^{(s)}$ does not dominate the total power in primordial gravitational waves $\mcP_h$.
For instance, with the parameters used in Fig.~\ref{three Phs}, 
the contribution of the vacuum fluctuation is $\mcP_h^{\rm (vac)}= 5.5\times 10^{-12}$ as shown as the black dotted line and hence the sourced contribution
$\mcP_h^{(s)}$ is subdominant for $p=1/2$ and $\ln(k/k_*)\lesssim 20$.
In such a case, the observed tensor tilt $\dd\ln(\mcP_{h}^{\rm (vac)}+\mcP_h^{(s)})/\dd\ln k$ is significantly suppressed compared to $n_T^{\rm power}$.
Of course, for a smaller $H$ and a larger $m_*$,
$\mcP_h^{(s)}$ becomes dominant and the vacuum contribution can be ignored.
Therefore, generally speaking, even if the observed tilt $n_T$ is not apparently large, the sourced contribution may be hidden under the vacuum fluctuation.
In principle, it might be feasible to discover the non-vacuum contribution to the tensor modes and thus access the axion potential at very high energy scales by comparing the two polarizations 
of gravitational waves. Furthermore, vacuum-generated gravitational waves are inherently highly gaussian, which is not true for sourced tensor modes \cite{Agrawal:2017awz}. Hence non-gaussianities can also be used to look for a non-vacuum gravitational wave component.  
Finally, tensor spectra with a blue tilt, or gaussian spectra peaked after the CMB-relevant scales have left the hozizon, could be observable using both CMB measurements as well as fall within the LISA band. This will provide a way  to probe the axion potential over a large field range $\Delta \chi$, since the CMB and LISA probe gravitational waves that are produced at different points during inflation.

\subsection{Type III modulated potential}

We now turn to the possibility of detecting the oscillating feature in the primordial gravitational wave spectrum $\mcP_h$ with a future observation of the B-mode polarization in the CMB.  
Based on the discussion in Section~\ref{sec:monodromy}, we use the following template of the GW power spectrum with a oscillatory feature,
\begin{equation}
\mcP_h^{(s)}(k)= \mcP_h^{(0)}\Big[1+A\sin \big(C\ln(k/k_*)+\theta \big)\Big],
\label{Ph template}
\end{equation}
where we have introduced four parameters:
\begin{enumerate}[label=(\roman*)]
\item the mean amplitude $\mcP_h^{(0)}$, 
\item the oscillation amplitude $A$, 
\item the frequency $C$ which corresponds to $\omega$ in Section \ref{sec:monodromy} and 
\item 
the constant phase $\theta$. 
\end{enumerate}
Since the scale $k_*$
is degenerate with the angle $\theta$ in Eq.~\eqref{Ph template}, we fix the scale $k_*=0.05\, {\rm Mpc}^{-1}$.
Although the observed tensor power spectrum $\mcP_h$ will in general be the sum of the intrinsic GW power $\mcP_h^{(vac)}$
coming from the vacuum fluctuation and the sourced one $\mcP_h^{(s)}$,
we ignore the former for simplicity by assuming $\mcP_h^{(vac)}\ll \mcP_h^{(s)}$.
The mean amplitude is fixed such that the corresponding tensor-to-scalar ratio is $10^{-2}$, namely $\mcP_h^{(0)}/\mcP_\zeta =0.01$.
Note that the template given in Eq.~\eqref{Ph template} becomes less accurate for a large modulation amplitude $A\gtrsim 0.3$ as we discussed in Section~\ref{sec:monodromy}.
Despite the fact that the GW template will deviate from the pure sinusoidal curve in such cases, $\mcP_h$ still exhibits
a periodic oscillation and Eq.~\eqref{Ph template} suffices for our purpose as a template with fewer parameters than Eq.~\eqref{analytic Ph}.

%
\begin{figure}[tbp]
  \begin{center}
  \includegraphics[width=120mm]{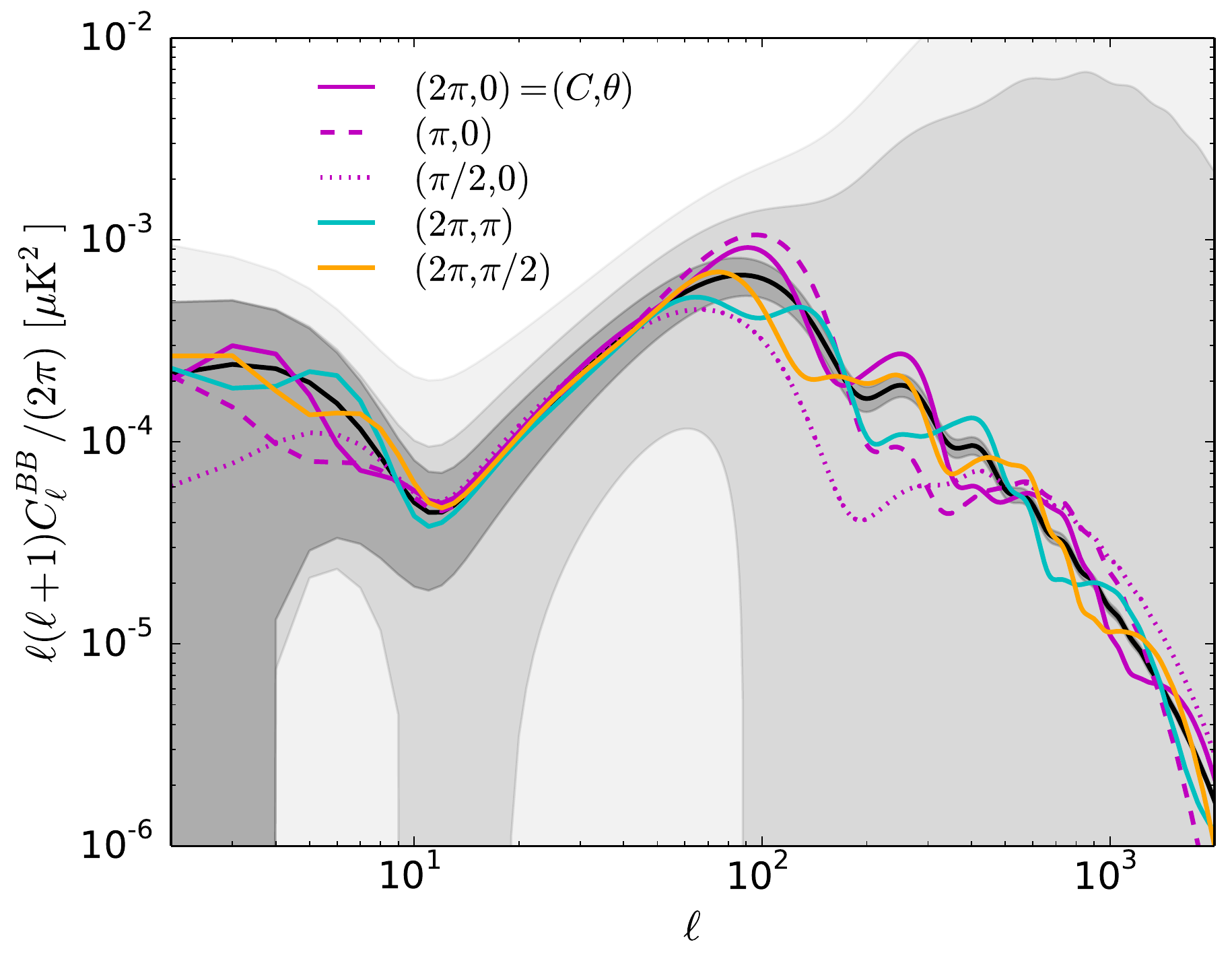}
  \end{center}
  \caption
      {CMB B-mode power spectra $C_\ell^{BB}$ for the template of Eq.~\eqref{Ph template}, corresponding to a modulated axion monodromy potential. The tensor to scalar ratio corresponding to the scale $k_* =0.05 \, \rm{Mpc}^{-1}$ is chosen as $r=0.01$. The various colored lines correspond to different choices of the modulation frequency $C$ and phase $\theta$, while the GW modulation amplitude is set to $A=0.9$ and the vacuum gravitational wave contribution is ignored. The black solid line shows the un-oscillatory case $\mcP_h(k)=\mcP_h^{(0)}$
for reference. The three-step shaded colors describe the regions within 2$\sigma$ deviations from the reference signal (black solid line) under three different assumptions for delensing and experimental uncertainties. We see that for the best-case scenario, i.e., a cosmic-variance-limited case (dark-shaded region), the modulation of the GW spectrum is in principle detectable.
}
 \label{ClBB9}
\end{figure}
%
%
\begin{figure}[tbp]
  \begin{center}
  \includegraphics[width=120mm]{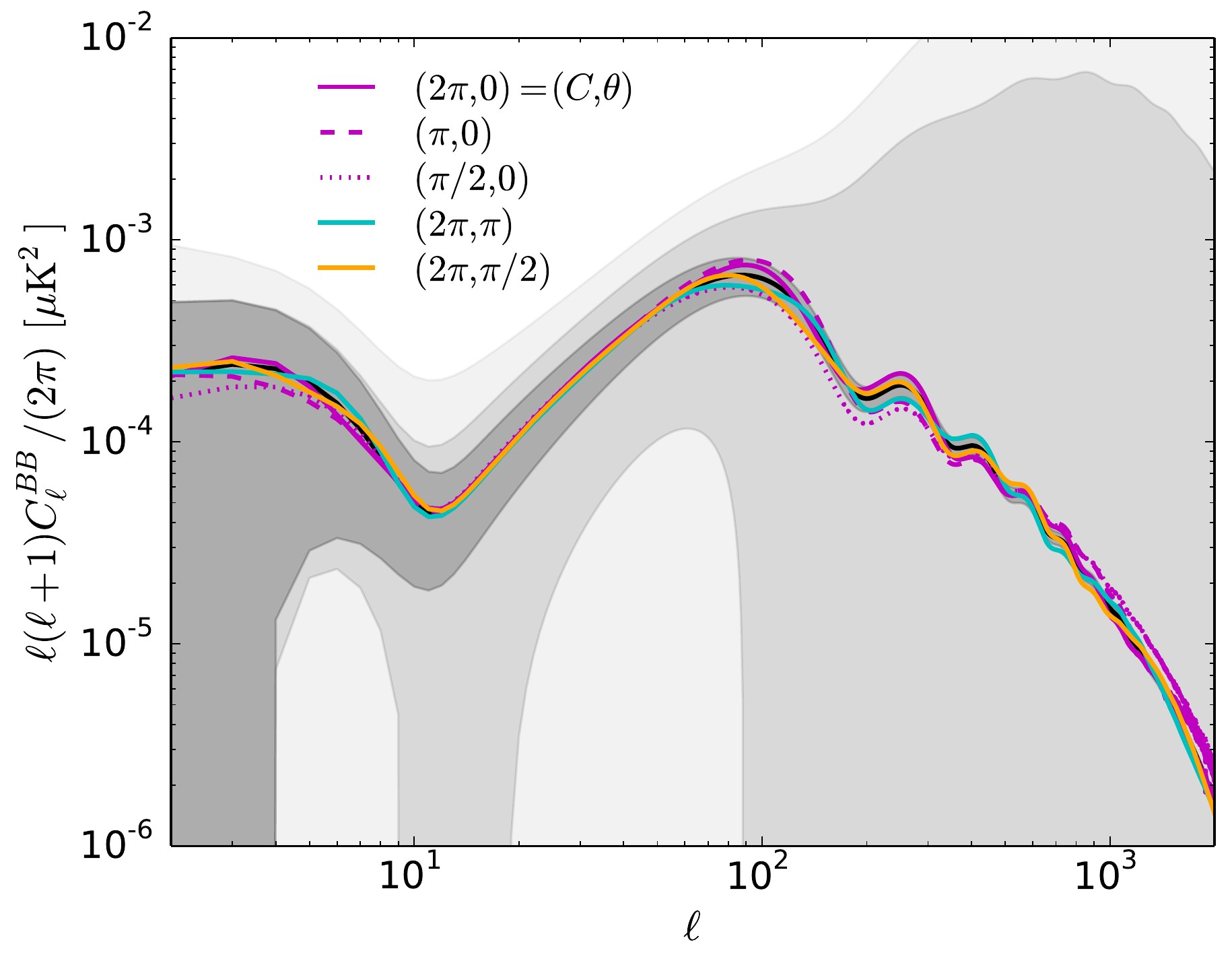}
  \end{center}
  \caption
  {CMB B-mode power spectra $C_\ell^{BB}$ for the template of Eq.~\eqref{Ph template} with $A = 0.3$ but otherwise the same parameters as Fig.~\ref{ClBB9}. The shaded regions unchange from Fig.~\ref{ClBB9}. We see that even for the best-case scenario (dark-shaded region), the modulation of the GW spectrum marginally detectable.
}
 \label{ClBB3}
\end{figure}
%
%
\begin{figure}[tbp]
  \begin{center}
  \includegraphics[width=120mm]{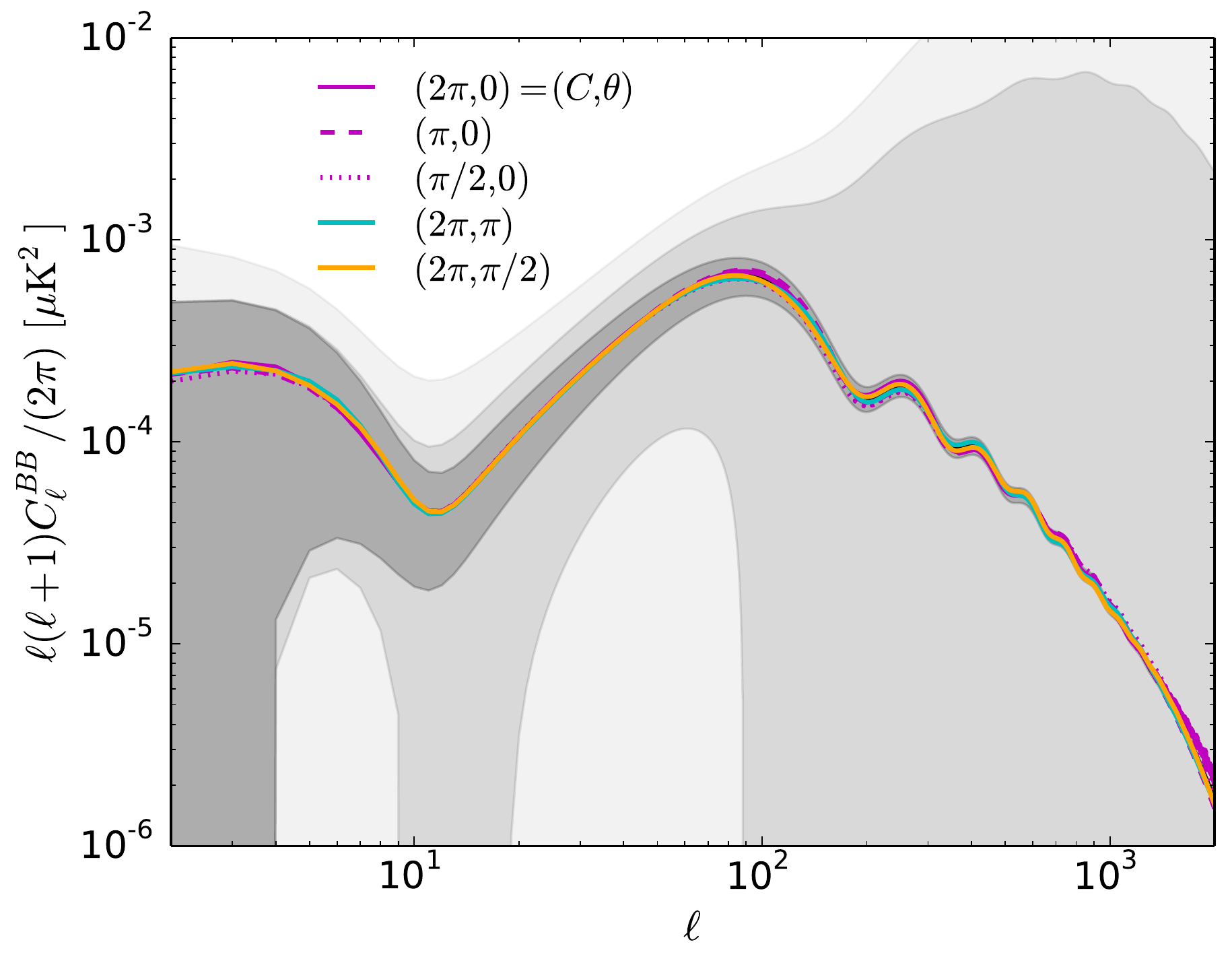}
  \end{center}
  \caption
      {CMB B-mode power spectra $C_\ell^{BB}$ for the template of Eq.~\eqref{Ph template} with $A = 0.1$ but otherwise the same parameters as Fig.~\ref{ClBB9}. The shaded regions unchange from Fig.~\ref{ClBB9}. We see that the modulation of the GW spectrum is too weak to be detectable, even for the best-case observational scenario.}
 \label{ClBB1}
\end{figure}
%
By varying the remaining three parameters $A,C,\theta$, we calculate the CMB angular power spectrum of the B-mode polarization $C_\ell^{BB}$ shown in Figs.~\ref{ClBB9}-\ref{ClBB1}. In particular, the oscillating amplitude $A$
is chosen as $A=0.9, 0.3, 0.1$ in Fig.~\ref{ClBB9}, Fig.~\ref{ClBB3} and Fig.~\ref{ClBB1},
respectively.
In these figures, the shaded regions represent the $2\sigma$ errors around the reference power spectrum for different experimental conditions. 
The darkest shade assumes perfect delensing and neglects experimental uncertainties coming from beams, inhomogeneous noises, masking and foreground contaminations and accordingly describes the level of cosmic variance. 
The grey region assumes no delensing and still neglects experimental uncertainties.
The lightest shade includes experimental uncertainties anticipated in LiteBIRD \cite{Matsumura:2016sri} without delensing. For the reference power spectrum, we adopt the mean amplitude $\mcP_h^{(0)}$, i.e., the case for $A = 0$ in Eq.~\eqref{Ph template}\footnote{This reference flat spectrum corresponds to the $p=1$ case of the power-law potential studied in Sec.~\ref{sec:powerlaw}
as well as the limit $\delta\to 0$ in Eq.~\eqref{eq:monodromydef}.
Alternatively, it can be also seen as the vacuum fluctuation $\mcP_h^{(vac)}$ for $r=10^{-2}$ without the sourced contribution $\mcP_h^{(s)}$.}, and it is shown as black solid lines in Figs.~\ref{ClBB9}-\ref{ClBB1}; thus, the departure from the shaded regions indicates high distinguishability between the oscillatory spectrum and the flat spectrum.

These figures imply that although it is challenging to detect the oscillating
feature of the primordial GWs by the upcoming B-mode observations, it can be observed in principle if the oscillating amplitude is large enough. In the three observational scenarios that we considered, GW modulations with $A\gtrsim 0.3$ can be detected if perfect --or sufficiently good-- delensing is attained and if the instrumental uncertainties are almost absent (ideally in the cosmic-variance-limited measurement.
It is found that an even larger $C>2\pi$ (faster oscillation) does not increase our detection capacity.
Notice that these results are robust with respect to changes in $r$ since resultant changes in the sourced GW amplitude are fully cancelled out by those in the sizes of errors (as both are proportional to $r$ in the cosmic-variance-limited case) and hence the signal to noise remains unchanged.

Due to the difficulty of detecting an inherent modulation in the GW power spectrum through the $C_\ell^{BB}$ alone, it will be useful or even necessary to combine CMB data with those of other experiments on smaller scales such as the GW interferometers and the pulser timing arrays~\cite{Obata:2016xcr}. Furthermore, other CMB statistics like
B-mode $3$-point correlators may be helpful to increase the signal-to-noise ratio and make the detection and mapping of the axion potential in case of GW production by an axion-$SU(2)$ sector feasible (See Ref.~\cite{Shiraishi:2016yun} for a relevant work on an axion-$U(1)$ model).


\section{Summary and Discussion}
\label{sec:summary}

In this work we have studied the production of gravitational waves (GW) which are sourced by the tensor perturbation of $SU(2)$ gauge field coupled to a spectator axion field. Our study went beyond the cosine type potential of the axion field $\chi$ that had been considered so far in the literature,
$V(\chi) \propto 1-\cos \left(\chi/f\right)$. 
String theory constructions provide a variety of potentials. Instead of considering them one by one, we categorized the axion potential into three main categories based on its features, that directly relate to the produced gravitational wave spectrum:
\begin{itemize}
\item Type I: A monotonic potential, convex or concave for the whole field-range of interest during inflation. A demonstration model of the type $U(\chi) \propto |\chi|^p$ was used.
\item Type II: A potential with a single inflection point $\chi_*$ during inflation, such that  $U'(\chi_*)=0$ and $U''(\chi_*)\ne 0 $. A generalized cosine potential  $U(\chi) \propto \left [ 1-\cos \left(\chi/f\right)\right ]^{p/2}$ was used as a demonstration model.
\item Type III: A monotonic potential with an oscillatory term. We examined the specific potential $U(\chi) \propto |\chi|+\delta\cos(\nu\chi)$ in detail, where $\delta$ and $\nu$ characterize the modulation amplitude and frequency.
\end{itemize}
We extended the conventional analysis on the background dynamics of this system and analytically solved the time evolution of the background axion and gauge fields. In particular, we developed a new perturbative treatment for modulated Type III potentials, since the oscillatory behavior of the background fields requires a prescription beyond the conventional approach under the slow-roll limit (e.g. \cite{Dimastrogiovanni:2016fuu}). In  previous works, the tensor perturbations are solved under the assumptions that the  gauge field background $Q$ and the velocity of the background axion field $\dot\chi$ are constant. In our case, however, the time evolution of the background fields, $Q(t)$ and $\dot{\chi}(t)$, caused by the new axion potentials is essential to predict the sourced GW power spectrum with new features. We scrutinized the GW power spectrum both analytically and numerically, underlining the limits of the analytical slow-roll results  and obtaining the following $k$-dependence of the sourced GW power spectra:
\begin{align}
&{\rm [Type\ I]}\quad~ \mcP_h^{\rm (s)}(k)\propto \left(\frac{k}{k_*}\right)^{n_T},
\qquad {\rm sign}(n_T) = {\rm sign}(p-1),
\\
&{\rm [Type\ II]}\quad \mcP_h^{(s)}(k) \propto
 \exp \left [- {{ \ln^2\left ( {k/ k_*}\right )\over 2\sigma_h^2} }\right ],
\\
&{\rm [Type\ II]}\quad \mcP_h^{(s)} \propto
 \Big(1+
{A} \sin(C \ln(k/k_*)+\theta)\Big), \quad ({A}\ll 1),
\end{align}
where $k_*$ is an arbitrary wave-number which can be identified with the CMB pivot scale $k_*=0.05 {\rm Mpc}^{-1}$.

The GW power spectrum of Type I potentials is well approximated by the standard power-law spectrum. It is interesting to note that the sign of the tensor tilt is solely determined by the axion potential power $p$. Thus, in this case, a scale-invariant GW spectrum can be realized for $p=1$, while steeper potentials $p>1$ lead to a red GW spectrum.
 The  consistency relation of single field inflation, $n_T=-r/8$ is generically violated and even a blue-tilted GW spectrum is possible for $p<1$.

 In the Type II case, where the inflaton crosses a single non-stationary inflection point of its potential during inflation, a universal prediction for the shape of the power spectrum is derived in the slow-roll approximation. The resulting GW spectrum exhibits a gaussian shape peaked around an arbitrary scale $k_*$, corresponding to the mode that exits the horizon around the time when the axion field crosses its inflection point. A faster-rolling axion leads to a more sharply peaked spectrum. The deviation of the spectrum from the gaussian away from the peak depends on the shape of the potential away from the inflection point, as well as the axion-gauge coupling and the resulting axion velocity $\dot \chi$.  
 
In the case of modulated axion monodromy potentials, denoted as Type III, the oscillatory form of the potential leads to a modulation in the GW spectrum. Both the amplitude of the GW modulation $A$ and the frequency $C$ of the GW spectrum in logarithmic wavenumber space depends on the axion monodromy potential parameters. It is interesting to note that the amplitude of the oscillatory part of the GW spectrum is suppressed for higher values of the modulation frequency $C/2\pi \gtrsim 0.1$. Furthermore, a large modulation amplitude $A\gtrsim 1$ results in a tensor template whose wavenumber dependence is more complicated than a pure sinusoidal modulation term. 
Computing the CMB angular power spectrum of the B-mode polarization $C_\ell^{BB}$ resulting from a modulated primordial GW spectrum and comparing it to proposed future experiments, such as LiteBIRD, we showed that it is in principle possible to 
distinguish the oscillatory signal from the scale-invariant fluctuation.
Such a detection requires a strong enough modulation parameter $A\gtrsim0.3$, as well as low 
experimental uncertainties and very good delensing.

Several intriguing possibilities arise by closely examining the different axion potential types. In the case of blue-tilted gravitational wave spectra arising from Type I potentials, the possibility  of detecting stochastic gravitational waves at LISA as well as the CMB can offer an unprecedented chance of studying axion potentials at high energies over a wide range of field values. This is also true in the case of Type II spectra peaked at scales much smaller than the CMB, which are   are also locally blue-tilted, when computed at  CMB scales. Chiral gravitational waves, like the ones produced in this class of models, that peak towards the end of inflation, can generate the observed matter-antimatter asymmetry through gravitational leptogenesis \cite{Alexander:2004us, Maleknejad:2014wsa, Maleknejad:2016dci, Caldwell:2017chz}. A connection of the gravitational wave signal to leptogenesis can not only probe the potential transversed by the spectator axion field throughout inflation, but also give hints on the reheat temperature and the nature of the neutrino sector, as discussed in Ref.~\cite{Adshead:2017znw}.


\acknowledgments
TF is in part supported by the Grant-in-Aid for JSPS Research Fellow No.~17J09103.
EIS acknowledges support by the NetherlandsÕ Organisation for Scientific Research (NWO) and the NetherlandsÕ Organization for Fundamental Research in Matter (FOM).
MS was supported by JSPS Grant-in-Aid for Research Activity Start-up Grant Number 17H07319. Numerical computations by MS were in part carried out on Cray XC50 at Center for Computational Astrophysics, National Astronomical Observatory of Japan.

\appendix
\section{Appendix: Numerical Calculation of Tensor Perturbations}
\label{Appendix: Numerical Calculation of Tensor Perturbations}

In this appendix, we describe the set up of the numerical computation for the tensor power spectrum. This involves solving a coupled system of equations for the tensor perturbations, which are the usual gravitational waves and the tensorial part of the $SU(2)$ gauge field, as defined in Eq.~\eqref{eq:deltaAmu}. 
 The quadratic action for the tensor degrees of freedom is written as~\cite{Agrawal:2017awz}
\begin{align}
S^{(2)} = \int &d\tau d^3 x  \Bigg[ 
\frac{1}{2}\psi'_{ij}\psi'_{ij}-\frac{1}{2}\partial_k\psi_{ij}\partial_k \psi_{ij}+\frac{1}{\tau^2}\psi_{ij}\psi_{ij}
\notag\\
&+\frac{1}{2}t'_{ij}t'_{ij} -\frac{1}{2}\partial_l t_{ij}\partial_l t_{ij}
-\frac{m_Q \xi}{\tau^2}t_{ij}t_{ij} +\frac{m_Q+\xi}{\tau}\epsilon^{ijk} t_{il}\partial_j t_{kl}
\notag\\
&+\frac{2\sqrt{\epsilon_E}}{\tau}\psi_{ij}t'_{ij}+\frac{2\sqrt{\epsilon_B}}{\tau}\psi_{jm}\epsilon_{aij}\partial_i t_{am} +\frac{2\sqrt{\epsilon_B}m_Q}{\tau^2}\psi_{ij}t_{ij} 
\Bigg],
\label{quadratic tensor action}
\end{align}
where $\psi_{ij}\equiv a\Mpl h_{ij}/2$, $t_{ij}$ is the traceless and transverse part of $\delta_{ai} \delta A^a_j$, and prime denotes the derivative with respect to conformal time.
We express both fields in Fourier space:
\begin{align}
\psi_{ij}(\tau,\bm x)&= \int \frac{\dd^3 k}{(2\pi)^3} e^{i \bm k \cdot \bm x} \left( e^{R}_{ij}(\hat{\bm{k}})\psi_R(\tau,\bm k)+ e^{L}_{ij}(\hat{\bm{k}})\psi_L(\tau,\bm k)\right),
\\
t_{ij}(\tau,\bm x)&= \int \frac{\dd^3 k}{(2\pi)^3} e^{i \bm k \cdot \bm x} \left( e^{R}_{ij}(\hat{\bm{k}})t_R(\tau,\bm k)+ e^{L}_{ij}(\hat{\bm{k}})t_L(\tau,\bm k)\right),
\label{mode decomposition}
\end{align}
%
where $e_{ij}^{L/R}(\hat{\bm{k}})$ is the left/right-handed polarization tensor which satisfies
\begin{equation}
e^{L}_{ij} (-\hat{\bm{k}})= e^{L*}_{ij} (\hat{\bm{k}})=e^{R}_{ij} (\hat{\bm{k}}),
\quad
i \epsilon_{ijk} k_i e_{jl}^{L/R}(\hat{\bm{k}})=
\pm k e_{kl}^{L/R}(\hat{\bm{k}}),
\quad
e^{L}_{ij} (\hat{\bm{z}}) =\frac{1}{2}\left(\begin{array}{ccc}
1 & i & 0 \\
i & -1 & 0 \\
0 & 0 & 0 \\
\end{array}\right).
\label{Pol property}
\end{equation}
Without loss of generality, we can assume $\dot{\chi}>0 $ leading to $ Q>0$ and $ m_Q>0$. With this choice the left-handed modes are not amplified and can be safely neglected.  We thus focus on the right-handed modes.
One can show that the quadratic action of the right-handed tensor modes is written as
\begin{equation}
S^{(2)}=\frac{1}{2}\int\dd\tau\frac{\dd^3 k}{(2\pi)^3}\left[
\Delta'^\dag \Delta'+ \Delta'^\dag K \Delta- \Delta^\dag K \Delta'
-\Delta^\dag \Omega^2 \Delta\right],
\end{equation}
with
\begin{align}
&\Delta=\begin{pmatrix}\psi_R \\
t_R \\
\end{pmatrix},
\quad
K=\frac{\sqrt{\epsilon_E}}{\tau}\begin{pmatrix}0 & -1 \\
1 & 0 \\
\end{pmatrix},
\notag\\
&\Omega^2 = \begin{pmatrix}k^2-\frac{2}{\tau^2} & -2\frac{\sqrt{\epsilon_B}m_Q}{\tau^2}-2\frac{\sqrt{\epsilon_B}k}{\tau}-\frac{\sqrt{\epsilon_E}}{\tau^2} \\
-2\frac{\sqrt{\epsilon_B}m_Q}{\tau^2}-2\frac{\sqrt{\epsilon_B}k}{\tau}-\frac{\sqrt{\epsilon_E}}{\tau^2}\quad & k^2+ \frac{2m_Q \xi}{\tau^2}+2\frac{m_Q+\xi}{\tau}k\\
\end{pmatrix}.
\\
&\epsilon_B \equiv  \frac{g^2 Q^4}{H^2 \Mpl^2},
\qquad 
\epsilon_E\equiv  \frac{(\dot{Q}+HQ)^2}{H^2 \Mpl^2}.
\end{align}
It is straightforward to obtain the equation of motion:
\begin{equation}
\Delta''+2K\Delta'+(\Omega^2 + K')\Delta=0.
\label{Matrix EoM}
\end{equation}
It should be noted that since this system has both kinetic mixing and mass mixing, it cannot be diagonalized. 
Hence we solve the evolution of four modes which are the intrinsic perturbations of $\psi_R$ and $t_R$ originating from the vacuum fluctuation and the induced perturbations of $\psi_R$ and $t_R$ sourced by the intrinsic perturbation of the other field through the mixing effect, as in Ref.~\cite{Dimastrogiovanni:2012ew}:
\begin{equation}
\hat{\Delta}=
\begin{pmatrix}\hat{\psi}_R(\tau,\bm k) \\
\hat{t}_R(\tau,\bm k) \\
\end{pmatrix}
=\begin{pmatrix}\Psi_k^{\rm int}(\tau) & \Psi_k^{\rm src}(\tau) \\
T_k^{\rm src}(\tau) & T_k^{\rm int}(\tau) \\
\end{pmatrix}
\begin{pmatrix} \hat{a}_{\bm k} \\
\hat{b}_{\bm k} \\
\end{pmatrix}+{\rm h.c.},
\label{Matrix quantization}
\end{equation}
where $\hat{a}_{\bm{k}}/\hat{a}_{\bm{k}}^\dag$ and $\hat{b}_{\bm{k}}/\hat{b}_{\bm{k}}^\dag$ are two independent sets of creation/annihilation operators.
Since $\psi$ and $t$ are decoupled in the sub-horizon limit, it is reasonable to assume that $\Psi_k^{\rm int}$ and $T_k^{\rm int}(\tau)$ are in the  Bunch-Davies vacuum in the distant past. However, the amplitudes of $\Psi_k^{\rm src}$ and $T_k^{\rm src}$ vanish at the initial time:
\begin{equation}
\begin{aligned}
\lim_{|k\tau|\to \infty}
\begin{pmatrix}\Psi_k^{\rm int}(\tau) & \Psi_k^{\rm src}(\tau) \\
T_k^{\rm src}(\tau) & T_k^{\rm int}(\tau) \\
\end{pmatrix}
&=\frac{1}{\sqrt{2k}}\begin{pmatrix}1 & 0 \\
0 & 1 \\
\end{pmatrix},
\\
\lim_{|k\tau|\to \infty} \frac{\dd}{\dd \tau}
\begin{pmatrix}\Psi_k^{\rm int}(\tau) & \Psi_k^{\rm src}(\tau) \\
T_k^{\rm src}(\tau) & T_k^{\rm int}(\tau) \\
\end{pmatrix}
&=-i\sqrt{\frac{k}{2}}\begin{pmatrix}1 & 0 \\
0 & 1 \\
\end{pmatrix},
\end{aligned}
\end{equation}
where we suppress arbitrary constant phases.
The equations of motion  are derived from Eq.~\eqref{Matrix EoM},
\begin{multline}
\begin{pmatrix}\partial_x^2 \Psi_k^{\rm int} & \partial_x^2\Psi_k^{\rm src} \\
\partial_x^2T_k^{\rm src} &\partial_x^2 T_k^{\rm int} \\
\end{pmatrix}
+2\frac{\sqrt{\epsilon_E}}{x}
\begin{pmatrix}0 & -1 \\
1 & 0 \\ \end{pmatrix}
\begin{pmatrix}\partial_x \Psi_k^{\rm int} & \partial_x\Psi_k^{\rm src} \\
\partial_xT_k^{\rm src} &\partial_x T_k^{\rm int} \\
\end{pmatrix}
\\+
\begin{pmatrix}1-\frac{2}{x^2} & -\frac{2\sqrt{\epsilon_B}m_Q}{x^2}+\frac{2\sqrt{\epsilon_B}}{x} \\
-\frac{2(\sqrt{\epsilon_B}m_Q+\sqrt{\epsilon_E})}{x^2}+\frac{2\sqrt{\epsilon_B}}{x}\quad & 1+ \frac{2m_Q \xi}{x^2}-2\frac{m_Q+\xi}{x}\\
\end{pmatrix}\begin{pmatrix} \Psi_k^{\rm int} & \Psi_k^{\rm src} \\
T_k^{\rm src} & T_k^{\rm int} \\
\end{pmatrix}=0,
\label{Component EoM}
\end{multline}
where we have introduced $x\equiv -k\tau$.
One can numerically solve the coupled system of equations for the tensor mode functions by using  the numerically obtained background quantities, $m_Q(t), \,\xi(t), \, \epsilon_B(t)$ and $\epsilon_E(t)$.
With these mode functions, the dimensionless power spectra of the intrinsic and sourced right-handed gravitational waves are written as
\begin{equation}
\mathcal{P}_{h_R}^{\rm int}(\tau,k)=\frac{2H^{2}k^3\tau^2}{\pi^2\Mpl^2}  |\Psi_k^{\rm int}(\tau)|^2,
\qquad
\mathcal{P}_{h_R}^{\rm (s)}(\tau,k)=\frac{2H^{2}k^3\tau^2}{\pi^2\Mpl^2}  |\Psi_k^{\rm src}(\tau)|^2.
\end{equation}
We are interested in the super-horizon limit, $|k\tau|\to 0$, of the sourced part of the power spectrum $\mathcal{P}_{h_R}^{\rm (s)}(\tau,k)$. We denote this as $\mcP_h^{\rm (s)}(k)$ and it is the main quantity that is computed throughout this work and plotted in the relevant figures
\begin{equation}
\mcP_h^{\rm (s)} = \lim_{|k\tau|\to 0}
\frac{2H^{2}k^3\tau^2}{\pi^2\Mpl^2}  |\Psi_k^{\rm src}(\tau)|^2.
\end{equation}
%




\end{document}